\documentstyle[seceq,supplement,epsf,wrapft]{ptptex}

\pubinfo{No. 133, 1999}  
\notypesetlogo  

\title{
\Large \bf Various Approaches to Cosmological Gravitational Lensing 
in Inhomogeneous Models}

\author{%
Kenji {\sc Tomita}$^{1,}$\footnote{
 E-mail address: tomita@yukawa.kyoto-u.ac.jp}, 
Premana {\sc Premadi}$^{2,}$\footnote{
 E-mail address: premadi@astr.tohoku.ac.jp} 
and Takahiro T. {\sc Nakamura}$^{3,}$\footnote{
 E-mail address: nakamura@utaphp2.phys.s.u-tokyo.ac.jp}
}
\inst{%
$^1$ Yukawa Institute for Theoretical Physics, 
Kyoto University, Kyoto 606-8502, Japan
\\
$^2$ Astronomical Institute, Tohoku University, Sendai 980-8578, Japan
\\
$^3$ Department of Physics, The University of Tokyo, 
Tokyo 113-0033, Japan
}
\recdate{%
February 10, 1999
}

\abst{%

Gravitational lensing of distant objects caused by gravitational
tidal forces from inhomogeneities in the universe is weak in most
cases, but it is noticed that it gives a great deal of information 
about the universe, especially regarding the distribution of dark 
matter. The statistical values of optical quantities such as  
convergence, amplification and shear have been derived by 
many people using various approaches, which include the linear
perturbational treatment in the weak limit and the nonlinear treatment 
considering small-scale matter distribution. 

In this review paper we compare the following three main approaches:
(a) the approach in the multi-lens-plane theory; (b) the approach due
to the direct integration method; and (c) the perturbational
approach. 
 
In the former two approaches inhomogeneous matter distributions
are produced in the CDM model using $N$-body simulations (the P$^3$M 
code and the tree-code, respectively). In (c) the power spectrum
corresponding to the CDM model is used for the large-scale matter
distribution.

}

\begin{document}

\maketitle

\section{Introduction}
The propagation of light from distant objects like galaxies and QSOs
is deflected by the gravitational tidal forces caused by inhomogeneous
matter distribution between the objects and us. The so-called lens
effect creates conspicuous images such as multiple QSOs and arcs in
clusters of galaxies, owing to special positional relations among sources,
lens objects and the observer, but in most cases it causes small
deformations and amplification of optical images, which result from
the superposition of deflections from many lens objects on
cosmological scales. This so-called weak lensing gives us valuable 
information on the structure and evolution of the universe, especially 
regarding the distribution of dark matter, 
not only on large scales, but also on small scales around galaxies.

The statistical behavior of optical quantites such as 
convergence, amplification and shear due to weak lensing
has been studied by many people since the pioneering papers
by Gunn,\cite{rf:gunna}\tocite{rf:gunnb}  Weinberg\cite{rf:weina} and
Blandford and Jaroszy\'nski.\cite{rf:bj} For the derivations of these
quantities there have 
been various approaches which consist of the multi-lens-plane method, 
direct integration methods solving the null-geodesic equation, 
optical scalar equations and 
the equation of geodesic deviation, and perturbative methods.

\noindent 1.1. {\it Multi-lens-plane method}

This is one of the various methods numerically simulating 
 light propagation in 
inhomogenous model universes, which were derived and developed by
Blandford and Narayan,\cite{rf:bn86}
Schneider and Weiss,\cite{rf:sw88a}\tocite{rf:sw88b} Jaroszy\'nski 
et al.,\cite{rf:jppg} Jaroszy\'nski.\cite{rf:jaro91a}\tocite{rf:jaro92b}     
In this method we first build a finite number of planes normal to light rays 
between an observer and a source, and the three-dimensional
distribution of galaxies and dark matter as lens objects is replaced by 
their two-dimensional distribution in the planes onto which the matter 
in each interval is projected.
The deflection of light rays resulting from each lens plane is
calculated using geometrical optics, and the statistical averages of
optical quantities for many-ray bundles are derived. \cite{SCH92}

To use this method we must produce the inhomogeneous distribution of
lens objects in each lens plane. In early works (Schneider and 
Weiss,\cite{rf:sw88b} Paczy\'nski and Wambsganss,\cite{rf:pw89} Lee and
Paczy\'nski.\cite{rf:lp90}) the random distribution of lens objects with 
an equal mass was assumed, and so the realistic large-scale structure
of the matter distribution was neglected. Jaroszy\'nski et 
al.\cite{rf:jppg} constructed an inhomogeneous model using an $N$-body 
simulation (the PM code), but in this model  large-scale structure was
considered, while small-scale structure corresponding to galaxies 
was neglected.
Jaroszy\'nski\cite{rf:jaro91a}\tocite{rf:jaro92b} developed other numerical
models of matter distribution using the Zeldovich approximation,
located galaxies (as lens objects) correponding to the matter density, 
and assumed for them the mass spectrum due to the Schechter luminosity
function and their morphological types randomly.

Wambganss et al.\cite{rf:wam95} \tocite{rf:wam98} generated models for
the evolution of large-scale structures by $N$-body simulation in
the CDM models, investigated their
influence on light propagation and statistical results such as
the frequency and separation angles of multiple QSOs and the image
deformation of high-redshift objects. In their works, galaxies as lens
objects were not taken into account.

Recently Premadi et al.\cite{rf:prem98} have improved the work of 
Jaroszy\'nski et
al. by adopting an $N$-body simulation (P$^3$M code) and
considering the spatial distribution of galaxies with the mass
spectrum and morphological types (due to the morphological
type-density relation). They adopted Jaroszy\'nski's assumption that the 
main lens objects are galaxies and the background
matter outside galaxies has smooth distribution with radii
$\approx 1 h^{-1}$ Mpc. In \S 2 the recent result of 
studies by Premadi et al.\cite{rf:prem98} is given.

\noindent 1.2. {\it Direct integration methods}

There are three types of direct integration methods in which the
optical scalar equation (derived by Sachs\cite{rf:sachs61}), the
equation of geodesic deviation and the null-geodesic equation are
solved.  Kantowski's approach belongs to the first type. He derived
the averaged optical scalar equation and solved it in an inhomogeneous 
model universe with swiss cheese structure.\cite{rf:kant69}
Dyer and Roeder\cite{rf:dr3} studied the behavior of the
amplification applying this approach to the case of a nonzero
cosmological constant. Recently, Kantowski et al.\cite{rf:kant95} have
analyzed the lens effect in the observed redshift-magnitude relation
for type Ia supernovae.
Watanabe and Tomita\cite{rf:wt90} studied the behavior of shear in
the ray bundles in various models in which the random distribution of 
particles simulating galaxies and clusters of galaxies was assumed
for simplicity. As an example of the second type of direct integration 
method (treating the equation of geodesic deviation)
we cite a recent work of Holz and Wald,\cite{rf:hw98} who studied
the behavior of amplification in an inhomogeneous model with a
swiss cheese-like structure. In their case not all systems consisting of a 
central galaxy and the surrounding empty region  have the
compensated mass distribution, in contrast to Kantowski's swiss cheese
system. They derived the probability distribution of amplification,
which was found to be non-Gaussian in accord with 
Dyer-Oattes\cite{rf:dyoa83} and examined the significance of the amplification 
on the observation of type Ia supernovae (Holz\cite{rf:holz98}).
A drawback of this approach may be that it treats only ray bundles with
infinitesimal separation angles.

The third type of direct integration method is described in the works
of Tomita and Watanabe.\cite{rf:tw89}\tocite{rf:tw90} They treated the
multiple deflection of light rays directly by solving the
null-geodesic equation and studied the statistical averages of image
deformation in inhomogeneous universe models. The evolution of models
was derived under the periodic condition using Aarseth's individual
particle method for $N$-body simulations, in which particles are
regarded as galaxies or clusters of galaxies with the initial power
spectrum $n = 0$.  
The particles (with the same mass and particle number $N = 1331$)  
are placed and move in a periodic box with comoving length $\approx
50$ Mpc, and light rays also propagate in the box under 
gravitational forces from particles in the box.  
In Tomita and Watanabe,\cite{rf:tw89} the influence of lensing on the
CMB anisotropy was analysed and a negative result was obtained,
contrary to expectations (Kashlinsky,\cite{rf:ka88}
Tomita,\cite{rf:tom88}\tocite{rf:tpr89} Sasaki.\cite{rf:sasaki89})
In the latter paper,\cite{rf:tw90} examples of strong image
deformations and their statistics were studied.  
Using the same approach, Tomita\cite{rf:tom98a}\tocite{rf:tom98c} has
recently studied the statistical behavior of optical scalars in
more realistic inhomogeneous models with the CDM spectrum ($n = 1$).
All particles are assumed to have the same galactic mass, while their
radii depend on the model used. 
In the first two of these papers,\cite{rf:tom98a}\tocite{rf:tom98b} 
the lens effect was
analyzed in flat and open models, respectively, and in the third
paper,\cite{rf:tom98c} the distribution of angular diameter distances
in these models was derived. 

As for the number density of main lens
objects, there is an important difference between the work of Premadi 
and that of Tomita. In the former, adopting Jaroszy\'nski's assumption
mentioned above, the density of main lens objects (galaxies) is found
to be $\Omega_g
\approx 0.02$. In Tomita's works, on the other hand, the density
($\Omega_L$) of main lens objects (called {\it compact lens
objects}) is $\Omega_L = 0.2$ for three models with $\Omega_0 = 1$
and $0.2$ 
and $\Omega_L = 0.4$ for a model with $\Omega_0 = 0.4$, which are much larger
than $\Omega_g$. This difference between these two works comes from
the different estimation for the role of non-galactic clouds as lenses. 
At present it is not clear whether Jaroszy\'nski's assumption is 
realistic or
not, and so the matter outside galaxies may not be so smooth and may
behave as discrete clouds with radii $r_{\rm cl} \approx 100$ kpc,
consisting of dark matter and the baryon fluid. In these clouds a 
baryon fluid has experienced no dissipated collapse and has less
central concentration than in galaxies, but as weak lens objects they
may play a role similar to galaxies. 

An outline of Tomita's recent works is given in \S 3, and the results of
additional analyses are also given in the case (satisfying Jaroszy\'nski's
assumption) that the density of lens
objects is $\Omega_L = \Omega_g$ and the remaining objects are
assumed to be smoother, by constraining their radii so as to be $500
h^{-1}$kpc. By comparing the two results in the case of $\Omega_L = 0.2$ 
and the case of $\Omega_L = \Omega_g$, we find how lensing on the 
small scale is
dominated by non-galactic lens objects, if they exist. 

\noindent 1.3. {\it Perturbative methods}

The method for solving optical scalar equations perturbatively with respect 
to a small perturbed expansion and shear has been introduced by 
Gunn\cite{rf:gunnb} and developed by Babul and Lee\cite{rf:bl91} and
Blandford et al.\cite{rf:bland91}.  Gravitational influences from 
large-scale structures are involved in the terms giving the Ricci and
Weyl focusing, through the density power spectrum or
the numerically simulated matter distribution. The angular correlation 
functions of optical scalars were derived in connection with the
two-point density correlation function, and higher correlations also were
studied by Villumsen,\cite{rf:vill96} Bernardeau\cite{rf:bern97} and
Nakamura.\cite{rf:nakam97}

The perturbative formulation for the image deformations has been
derived independently from the null-geodesic equation
(Linder,\cite{rf:lind90}  Martinez et al.,\cite{rf:mart92} Martinez and 
Sanz\cite{rf:mart97}). This approach 
also has been applied to studies concerning weak image deformation of
high-redshift objects and the smoothing of the CMB anisotropy
(Cayon et al.,\cite{rf:cay93} Seljak\cite{rf:selj96} and Jain and 
Seljak\cite{rf:jain97}). In \S 4, a recent result obtained using the power
spectrum approach by Nakamura is given.  

\section{Lensing in the multi-lens-plane method}
\subsection{Methodology}

The inhomogeneities that perturb a light bundle as it travels from
a source to an observer are of sizes much smaller than the
distance that the light traverses.  This has motivated a number of
authors to use the multi-plane gravitational lensing
approach to study how light propagates in inhomogeneous universes
(e.g. Schneider and Weiss,\cite{rf:sw88a}\tocite{rf:sw88b}
Jaroszy\'nski et al.,\cite{rf:jppg} 
Jaroszy\'nski,\cite{rf:jaro91a}\tocite{rf:jaro92b} Wambsganss, Cen
and Ostriker\cite{rf:wam98}).
In this approach one first idealizes the inhomogeneities as being 
distributed on a series of thin lens planes which are arranged 
perpendicular to the line of sight. Then one assumes that lensing
only occurs in each of those planes. This way one can analyze the 
lensing properties of each plane separately and let the light beam
carry the lensing effect of each plane while propagating from one
plane to the next. For this paper to be self-contained
we give a brief review of multi-plane lensing theory, following closely
the description and notation of Schneider et al.\cite{SCH92}

\subsection{Multi-plane gravitational lensing}

\def\mib#1{\mbox{\boldmath $#1$}}

Consider $N$ lens planes located at redshifts $z_{i}$,
with $i=1,N$, and ordered such that $z_{i}<z_{j}$ for $i<j$. 
Figure~1 displays an example with $N=2$. All angles are
exaggerated for clarity. Each lens plane is characterized by its
respective surface mass density $\sigma_{i}(\mib{\xi}_{i})$,
where $\mib{\xi}_{i}$ is the impact vector of the ray in the 
$i$-th lens plane. Let $\hat{\mib{\alpha}}_i(\mib{\xi}_{i})$ 
denote the
deflection angle the light ray experiences in the $i$-th plane at a
position $\mib{\xi}_{i}$. {From} this geometry, we can derive the 
{\it lens equation},
\begin{equation}\label{p1}
\mib{\eta} = \frac{D_S}{D_1}\mib{\xi}_1 - \sum_{i=1}^{N} D_{iS}
            \hat{\mib{\alpha}}_i(\mib{\xi}_{i}) \,,
\end{equation}

\noindent where $\mib{\eta}$ is the
source position vector (in the source plane),
$\mib{\xi}_{i}$ is the impact vector in the $i$-th plane, 
$D_j$ is the angular diameter distance between
the $j$-th plane and the observer, and $D_{ij}$ is the angular diameter
distance between the $i$-th and $j$-th planes, with $S\equiv N+1$ 
identifying the source plane.
Knowing the impact vector $\mib{\xi}_1$ in the image plane,
the impact vector in subsequent planes can be 
obtained recursively using
\begin{equation}\label{p2}
\mib{\xi}_{j} = \frac{D_j}{D_1}\mib{\xi}_1 - \sum_{i=1}^{j-1} D_{ij} 
            \hat{\mib{\alpha}}_i(\mib{\xi}_{i}) \,.
\end{equation}

\noindent
The deflection angle is related to the surface density by
\begin{equation}\label{p3}
\hat{\mib{\alpha}}_i(\mib{\xi})=\frac{4G}{c^2}\int\!\!\!\int
		\sigma_i(\mib{\xi}')
            \frac{\mib{\xi}_i-\mib{\xi}'}{|\mib{\xi}_i-\mib{\xi}'|^2}d^2\xi'\,,
\end{equation}

\noindent
where $G$ is the gravitational constant, $c$ is the speed of light,
and the integral extends over the lens plane. We can rewrite this expression
conveniently as
\begin{eqnarray}
\label{p4}
\hat{\mib{\alpha}}_i(\mib{\xi}_i)&=&\nabla\hat\psi_i(\mib{\xi}_i)\,,\\
\label{p5}
\hat\psi_i(\mib{\xi}_i)&=&\frac{4G}{c^2}\int\!\!\!\int
         \sigma_{i}(\mib{\xi}')\ln |\mib{\xi}_i-\mib{\xi}'|d^2\xi'\,.
\end{eqnarray}

\noindent It is useful to rewrite these equations
in a dimensionless form. Define
for each lens plane a critical surface density as
\begin{equation}\label{p6}
\sigma_{i,\rm cr}=\frac{c^2D_S}{4\pi GD_iD_{iS}}\,,
\end{equation}

\begin{figure}
\epsfxsize=12.cm
\centerline{\epsfbox{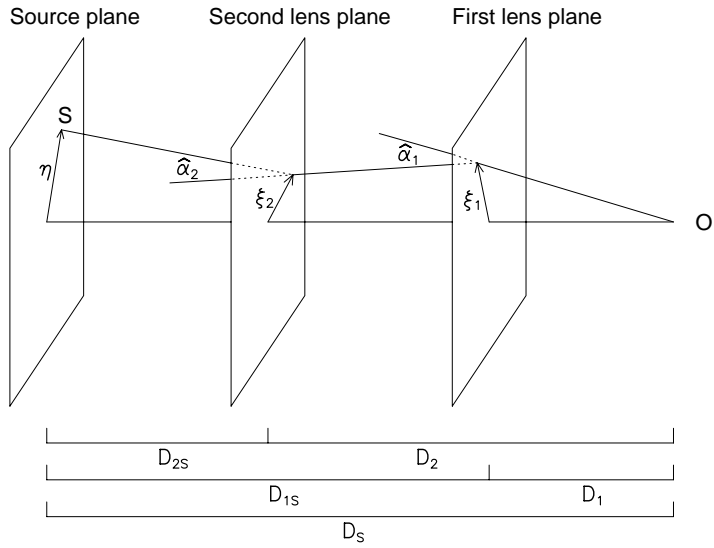}}
\caption{Multi-lens-planes.}
\label{fig:1}
\end{figure}

\noindent
and introduce the following dimensionless quantities :
\begin{eqnarray}
\label{p7}
{\bf \mib x}_{i}&=&\frac{\mib{\xi}_i}{D_i}\,,\qquad 1 \leq i \leq N+1 \,;\\
\label{p8} \kappa_{i}({\bf \mib x}_i)
&=&\frac{\sigma_i}{\sigma_{i,\rm cr}}\,,\qquad 1 \leq i \leq N \,.
\end{eqnarray}

\noindent Equations (\ref{p2}), (\ref{p4}) and (\ref{p5}) reduce to
\begin{eqnarray}
\label{p9}
{\bf\mib x}_{j} &=& {\bf\mib x}_{1} - \sum_{i=1}^{j-1} \beta_{ij}
                \mib{\alpha}_{i}({\bf\mib x}_{i})\,,\\
\label{p10}
\mib{\alpha}_{i}({\bf\mib x}_i) &=& \nabla \psi_{i}({\bf\mib x}_i)\,,\\
\label{p11}
\psi_{i}({\bf\mib x}_{i})&=& \frac{1}{\pi}\int\!\!\!\int
         \kappa_{i}({\bf\mib x}') \; \ln |{\bf\mib x}_i-{\bf\mib x}'|d^2x'\,,
\end{eqnarray}

\noindent where
\begin{equation}\label{p12}
\beta_{ij}=\frac{D_{ij}D_S}{D_jD_{iS}}\,,
\end{equation}

\noindent 
and the gradient is now taken relative to ${\bf\mib x}_i$. 
We can invert equation~(\ref{p11}) and obtain
\begin{equation}\label{p13}
\nabla^{2}\psi_i=2\kappa_i\,.
\end{equation}

\noindent
To compute the scaled position ${\bf\mib y}\equiv{\bf\mib x}_S$ of the source
in the source plane, we simply set $j=N+1$. Equation~(\ref{p12}) gives
$\beta_{iS}=1$, and Eq.~(\ref{p9}) becomes
\begin{equation}\label{p14}
{\bf\mib y} \equiv {\bf\mib x}_{N+1} = {\bf\mib x}_{1} - \sum_{i=1}^{N} 
                                \mib{\alpha}_{i}({\bf\mib x}_{i})\,.
\end{equation}

\noindent
This ray-tracing equation is a mapping from the image plane 
($i=1$) onto the source plane ($i=N+1$).

The effect of each lens plane on the evolution of the beam 
is described by the Jacobian matrix
\begin{equation}\label{p15}
{\bf \mib A}_i({\bf \mib x}_i)
=\left(\matrix{1-\psi_{i,11} &  -\psi_{i,12} \cr
                -\psi_{i,21} & 1-\psi_{i,22} \cr}\right)\,,
\end{equation}

\noindent
where the commas denote differentiation
with respect to the components of ${\bf \mib x}_i$.
Since $\psi_{i,12}=\psi_{i,21}$, and Eq.~(\ref{p13}) gives 
$\psi_{i,11}+\psi_{i,22}=2\kappa_i$, we can rewrite Eq.~(\ref{p15}) as
\begin{equation}\label{p16}
{\bf \mib A}_i=\left(\matrix{1-\kappa_i-S_{11} & -S_{12}\cr
-S_{12} & 1-\kappa_i+S_{11} \cr }\right)\,,
\end{equation}

\noindent where
\begin{eqnarray}
\label{p17}
S_{11}&=&\frac{1}{2}(\psi_{i,11}-\psi_{i,22})\,,\\
\label{p18}
S_{12}&=&\psi_{i,12}=\psi_{i,21}\,.
\end{eqnarray}

\noindent We now define
\begin{equation}\label{p19}
S_i=(S_{11}^2+S_{12}^2)^{1/2}\,.
\end{equation}

\noindent The determinant and trace of 
${\bf \mib A}_i$ can be
expressed entirely in terms of $\kappa_i$ and $S_i$, as follows:
\begin{eqnarray}
\label{p20}
\det\,{\bf \mib A}_i&=&(1-\kappa_i)^{2}-S_i^{2}\,,\\
\label{p21}
{\rm tr}\,{\bf \mib A}_i&=&2(1-\kappa_i)\,.
\end{eqnarray}

\noindent The quantities $\mu_i\equiv1/(\det{\bf \mib A}_i)$, 
$1-\kappa_i$, and 
$S_i$ are called the {\it magnification}, {\it convergence} 
(or Ricci focusing), and {\it shear}, respectively.

To compute the cumulative effect of all the lens planes, we consider
the Jacobian matrix of the mapping given by Eq.~(\ref{p14}),
\begin{equation}\label{p22}
{\bf\mib B}({\bf\mib x})=\frac{\partial{\bf\mib y}}{\partial{\bf\mib x}_1}
={\bf\mib I}-\sum_{i=1}^N\frac{\partial\mib{\alpha}_i}{\partial{\bf\mib x}_1}
={\bf\mib I}-\sum_{i=1}^N{\bf\mib U}_i{\bf\mib B}_i\,,
\end{equation}

\noindent where $\bf\mib I$ is the $2\times2$ identity matrix,
and ${\bf\mib U}_i$ and ${\bf\mib B}_i$ are defined by
\begin{eqnarray}
\label{p23}
{\bf\mib U}_i&=&\frac{\partial\mib{\alpha}_i}{\partial{\bf\mib x}_i}\,,\\
\label{p24}
{\bf\mib B}_i&=&\frac{\partial{\bf\mib x}_i}{\partial{\bf\mib x}_1}\,.
\end{eqnarray}

\noindent After substituting Eq.~(\ref{p10}) into Eq.~(\ref{p23}), we
obtain
\begin{equation}\label{p25}
{\bf\mib U}_i={\bf\mib I}-{\bf\mib A}_i=
\biggl(\matrix{ \psi_{i,11} & \psi_{i,12} \cr
                \psi_{i,21} & \psi_{i,22} \cr }\biggr)
\,,
\end{equation}

\noindent where ${\bf\mib A}_i$ is given by Eq.~(\ref{p15}).
Hence, ${\bf\mib U}_i$ describes
the effect the $i$-th plane would have on the beam if all
the other planes were absent, and Eq.~(\ref{p22}) simply combines the
effect of all the planes. To compute the matrices ${\bf\mib B}_i$,
we differentiate Eq.~(\ref{p9}) and get
\begin{equation}\label{p26}
{\bf\mib B}_j={\bf\mib I}-\sum_{i=1}^{j-1}\beta_{ij}{\bf\mib U}_i
{\bf\mib B}_i\,.
\end{equation}

\noindent Since ${\bf\mib B}_1={\bf\mib I}$, we can use Eq.~(\ref{p26}) 
to compute all matricies ${\bf\mib B}_i$ by recurrence.

\subsection{The numerical algorithm}

We use the P$^3$M algorithm (Hockney and Eastwood \cite{rf:HE81}) for 
the $N$-body
simulations of the large scale structure (LSS) of the universe. The 
calculations evolve a system of gravitationally interacting particles in
a cubic volume with triply periodic boundary conditions, comoving with the
Hubble flow. The forces on the particles are computed by solving the
Poisson equation on a cubic grid using a Fast Fourier Transform method.

\subsubsection{Building the lens planes}
To implement the multi-plane lens method, we divide the space between
the source and the observer into a chain of cubic boxes of equal 
comoving size, L$_{\rm box}$. We first need to determine the
redshifts of the interfaces between these cubic boxes. Let us assume that
the photons that reach the observer at present entered a particular
box at time $t'$ and redshift $z'$ and exited that box at time $t$ and
redshift $z$. The redshifts $z'$ and $z$ are related by (Premadi et
al.\cite{rf:prem98})
\begin{equation}\label{p27}
L_{\rm box}=\int_{t'}^{t}\big[1+z(t)\big]c\,dt\,.
\end{equation}
Using this equation, with the appropriate relation for $z(t)$,
we can find the redshifts of the interfaces. The front
side of the box closest to the the observer is, by definition, 
at $z=0$. Plugging
this value into Eq.~(\ref{p27}) gives us the redshift $z'$ of
the back side of the box, which is also the redshift $z$ of the front side
of the next box. Then, by using Eq.~(\ref{p27}) recursively, we can 
compute the redshifts of all the interfaces. 

As the structures inside the box might evolve while the light beam travels
across it, we decide the plane onto which we project the mass 
distribution 
to be at redshift where the density contrast is equal to the
time-averaged density contrast of the box.

\subsubsection{The galaxy distribution}
We consider the LSS at present ($z=0$) resulting from
the P$^3$M simulations, and we design an empirical Monte-Carlo method
for locating galaxies in the computational volume, based on the
constraints that (1) galaxies should be predominantly located in the
densest regions, and (2) the resulting distribution of galaxies
should resemble the observed distribution on the sky.
Our method is the following. We divide the present
computational volume into $128^3$ cubic cells of size $1\,{\rm Mpc}^3$, and
compute the matter density $\rho$ at the center of each 
cell using the same mass assignment scheme
as in the P$^3$M code. We then choose a particular density
threshold $\rho_{\rm t}$. We locate $N$ galaxies in each cell, where
$N$ is given by
\begin{equation}
N={\rm int}\biggl(\frac{\rho}{\rho_{\rm t}}\biggr)\,.
\end{equation}

\noindent The actual location of each galaxy is chosen to be
the center of the cell, plus a random offset that is of the order 
of the cell size.
This eliminates any spurious effect introduced by the use of a grid.
We then experiment with various values of the density
threshold $\rho_{\rm t}$ until the total number of galaxies comes out to be
of order 40000. This gives a number density of 
$\sim0.02\,{\rm galaxies}/{\rm Mpc}^3$. This method bears some
similarities with that used by 
Jaroszy\'nski.\cite{rf:jaro91a}\tocite{rf:jaro92b}
Tests showed that the observed galaxy 2-point correlation function
is fairly well reproduced (Martel et al.\cite{rf:MPM98}).

The morphological type of each galaxy is determined by using a 
combination of the known relations between the present distribution 
of morphological
types and the surface density of galaxies along with a Monte-Carlo method.

The locations of each galaxy at higher redshifts are determined by
combining the distribution of the galaxies and that of the particles from
the P$^3$M simulation, i.e., by following the position of the nearest particle.

We adopt the galaxy models described in Jaroszy\'nski. 
\cite{rf:jaro91a}\tocite{rf:jaro92b}
The projected surface density of each galaxy is given by
\begin{equation}
\sigma(r) = \cases{\displaystyle
            \frac{v^2}{4G(r^2+r_c^2)^{1/2}}\,,  & $r<r_{\max}$\,; \cr
             0\,,                             & $r>r_{\max}$\, , \cr}
\end{equation}

\noindent where $r$ is the projected distance from the center.
The parameters $r_c$, $r_{\max}$ and $v$ are the core radius, maximum radius,
and the rotational velocity, respectively. These parameters are 
functions of the
luminosity and morphological types of each galaxy. We assume that the 
present galaxy luminosities are given by the Schechter luminosity function, 
\begin{equation}
n(L)dL=\frac{n_*}{ L_*}\left(\frac{L}{L_*}\right)^{\alpha}e^{-L/L_*}dL\,,
\end{equation}
\noindent
where $n(L)$ is the number density of galaxies per unit luminosity,
$\alpha=-1.10$, $n_*=0.0156h^{3}{\rm Mpc}^{-3}$, and $L_{B*}=1.3\times
10^{10}h^{-2}L_{\odot}$, where $L_B$ is the luminosity
in the $B$ band (Efstathiou et al. \cite{rf:EEP88}). For each galaxy
we generate a luminosity $L$ according to this distribution, and we combine
it with the galaxy morphological type to determine the values of the
parameters $v$, $r_c$ and $r_{\rm max}$ for that galaxy.

\subsection{The experiments}
\subsubsection{The cosmological models}
We are currently appyling the algorithm described above to perform a large
cosmological parameter survey. We focus on the CDM model whose
density fluctuation power spectrum is described in detail in Bunn and White.
\cite{rf:BW97}
This power spectrum is characterized by 6 independent parameters :
(1) the density parameter $\Omega_{0}$; (2) the contribution $\Omega_B$ of 
the baryonic matter to the density parameter; (3) the cosmological constant
$\lambda_0$; (4) the Hubble constant $H_0$; (5) the temperature
$T_{\rm CMB}$
of the cosmic microwave background radiation; and (6) the tilt $n$ of
the power spectrum. We initially set $T_{\rm CMB}=2.7 K$ and 
$\Omega_B = 0.015h^{-2}$, thus reducing the dimensionality of the 
parameter space to 4. The normalization of the power spectrum is often 
described in terms of the rms density fluctuation $\sigma_8$ at a
scale of $8h^{-1}$Mpc, which is a function of the 6 aforementioned
parameters. We invert this relation, treating $\sigma_8$ as an 
independent parameter, and the tilt $n$ as the dependent parameter.
Presently we have 43 models with various (but restricted) values of
the above first 4 parameters. 

\subsubsection{The ray shooting}
In each model we propagate beams of $341^2$ light rays backward to 
redshift $z_s \approx 3$. The beam has a size of 21.9 arcseconds,
and the spacing between rays  is 0.064 arcseconds. To analyze the
results of the experiments, we lay down grid on the source plane,
divided into 31 x 31 cells. Cells which have more than the average
number of rays ($341^2/31^2$=121) indicate sources which are magnified.
The fraction of such cells gives us the magnification probability.
By taking the rays located in such cells, and tracing them back to the
image plane, we can study the shape of the images, and in particular,
compute the fraction of the magnified sources that have double images.
Figure~2 displays a few examples of images of a circular source.

\begin{figure}
\epsfxsize=12.cm
\centerline{\epsfbox{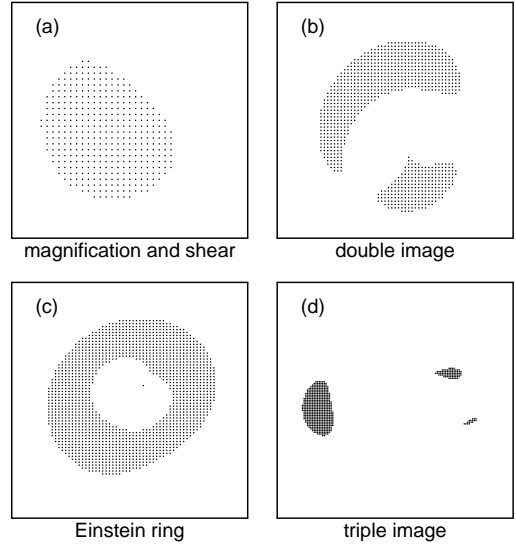}}
\caption{Various images of circular sources.}
\label{fig:2}
\end{figure}

\subsection{Preliminary results}
We have so far conducted between 30 and 60 ray shooting rays on all of the
models, but results of only 35 models have been analyzed. 
In a previous work (Premadi et al. 1998) we tested the methodology
we have described here by studying the statistics of magnification and
shear with respect to the lens redshift for a given source redshift.
Among other things, we have found that the lensing effect is most prominent at 
intermediate redshifts although the structures are more evolved at
lower redshifts. The redshift where most lensing occurs depends upon
the cosmological models. 
We also show that the magnification is dominated by convergence, with shear
contributing less than one part in \rm{$10^4$}.

In the current work we study the magnification
probability, $P_m$, the probability of double-image events given a lower limit
of magnification, $P_2$, and the distribution of the separation angle in the
double image events.

\begin{figure}
\epsfxsize=18.cm
\centerline{\epsfbox{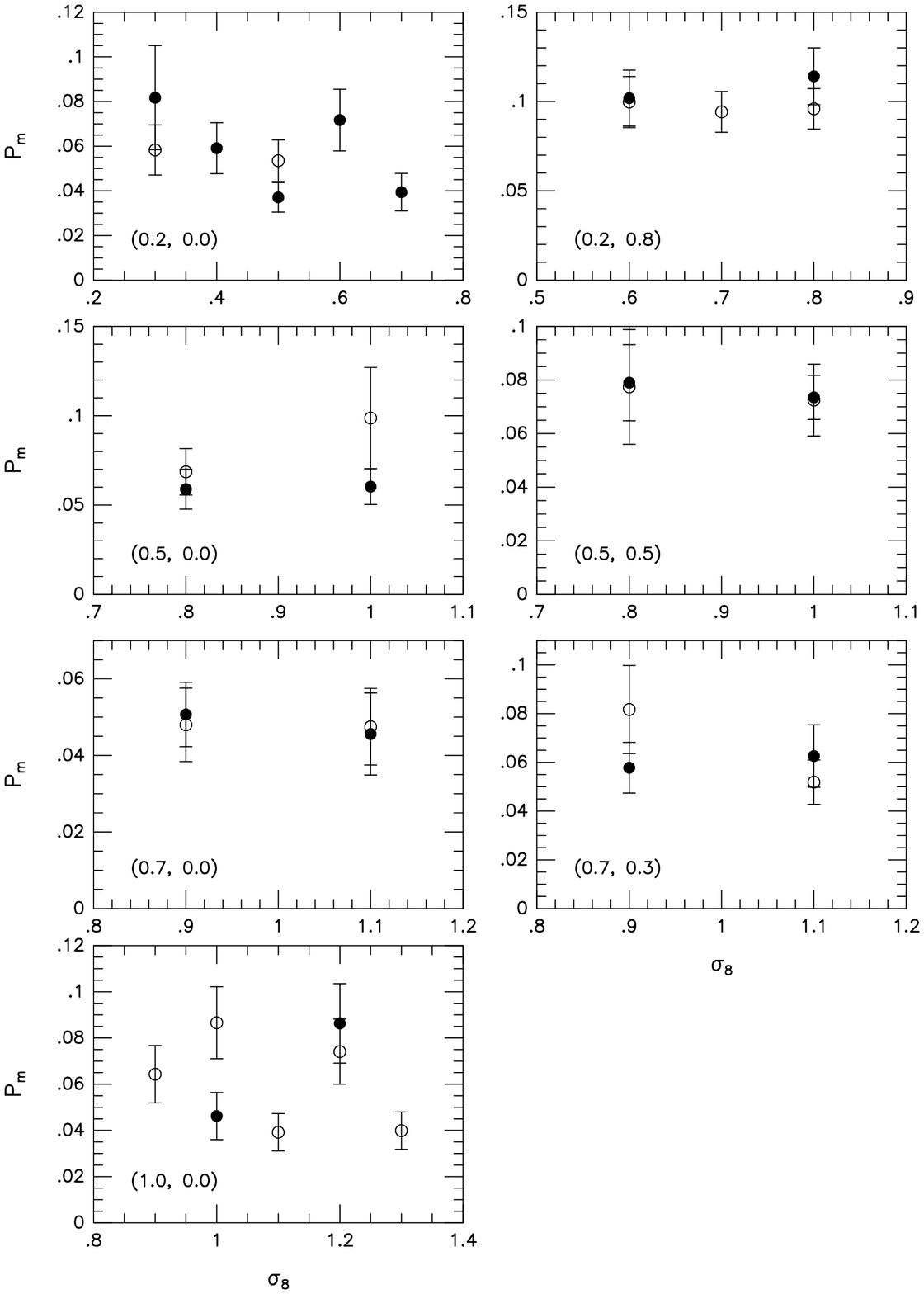}}
\caption{Magnification probability $P_m$ of magnification 1.2 and larger
as a function of the rms density fluctuation
$\sigma_8$. The numbers in parentheses indicate the values of 
($\Omega_0,\lambda_0$). Open circles correspond to $H_0=65$ and  
filled circles to $H_0=75$.}
\label{fig:3}
\end{figure}
\begin{figure}
\epsfxsize=18.cm
\caption{Multi-image probability $P_2$ as a function of magnification 
probability $P_m$. The numbers in parentheses indicate the values of 
($\Omega_0,\lambda_0$). Open circles correspond to $H_0=65$ and
circles to $H_0=75$.}
\centerline{\epsfbox{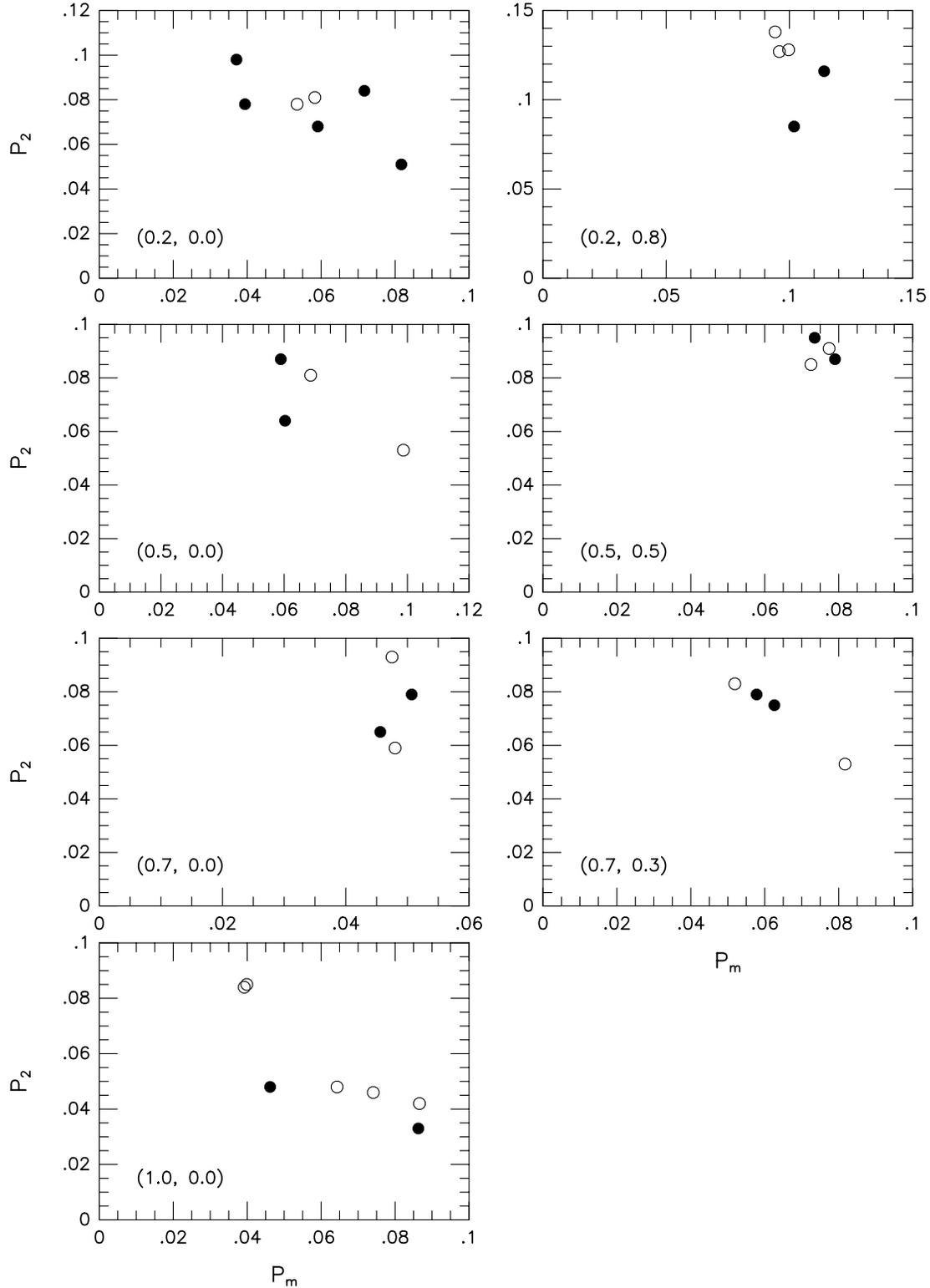}}
\label{fig:4}
\end{figure}
\begin{figure}
\epsfxsize=20.cm
\caption{The distribution of the image separation in arc seconds. The counts
are not normalized to the numbers of runs which are not the same for all
models. Top row displays results for models with $\Omega_0=1.0$,
$\lambda_0=0.0$. The middle row for models with $\Omega_0=0.2$,
$\lambda_0=0.0$, and the bottom row for models with $\Omega_0=0.2$, 
$\lambda_0=0.8$.}
\centerline{\epsfbox{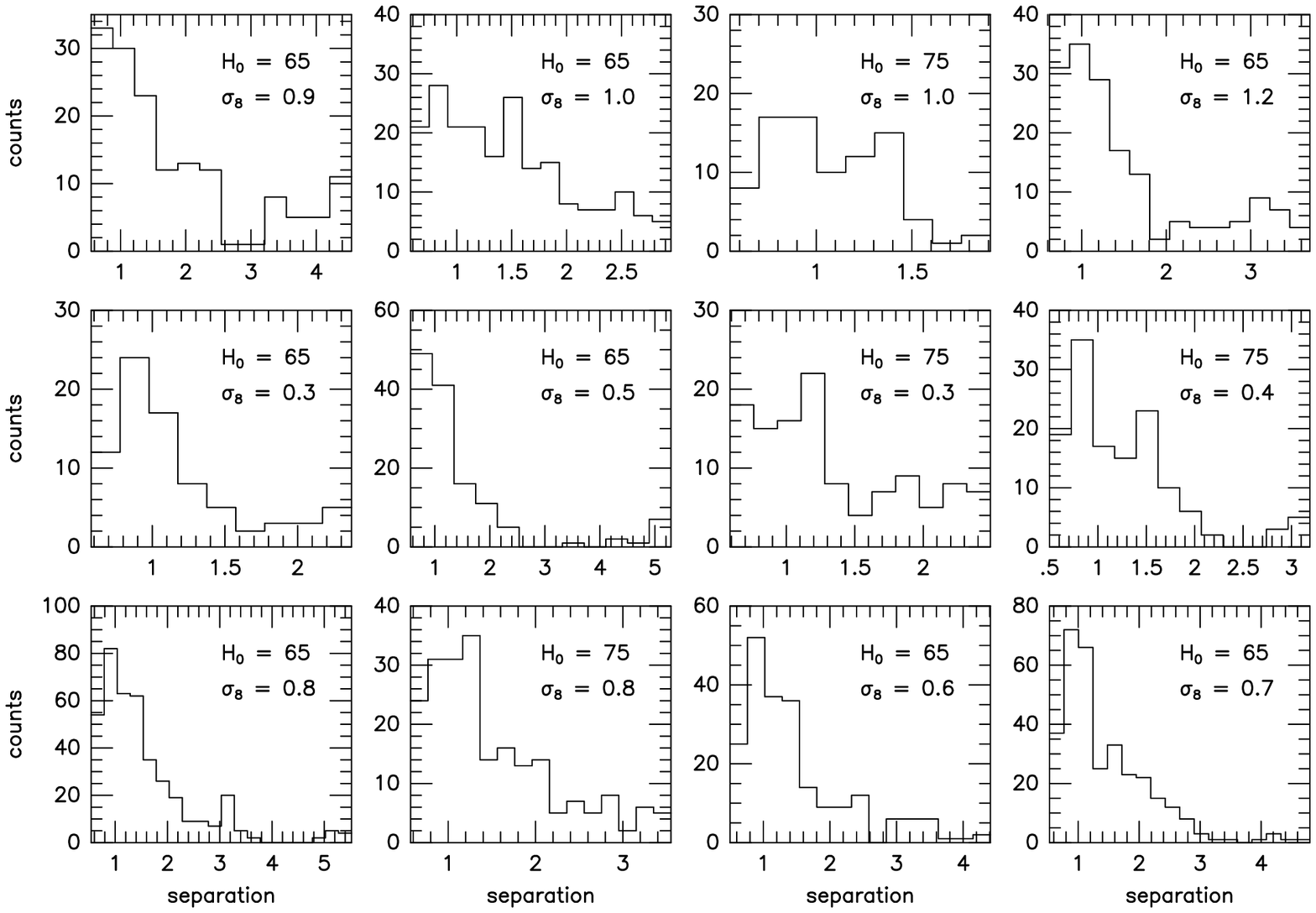}}
\label{fig:5}
\end{figure}


In Fig. 3, despite the still insufficient statistics, for the majority
of models, we observe a tendency for $P_m$ to decrease with increasing
$\sigma_8$, and for the cases in which we do not observe this tendency, it is
not ruled out because of the large error bars. Having an anticorrelation
between $P_m$ and $\sigma_8$ is seemingly counter-intuitive, since
$\sigma_8$ measures the amplitude of the density fluctuation, which are
responsible for lensing. The explanation is that the magnification is
caused primarily by the matter located near the beam, whereas matter located
far from the beam are responsible for the shear. In cosmological models
like CDM, structure formation proceeds hierarchically. Small structures
form first, then merge to form larger structures. A larger $\sigma_8$
implies that this hierarchical merging process is more advanced. This means
that the clusters are more massive, and possibly also denser, but there
are fewer of them, thereby enlarging voids between them. The light beam
is therefore less likely to hit or pass near any cluster, resulting in 
a smaller magnification probability for larger $\sigma_8$. However, if the
beam does hit a cluster, we would expect the effect to be stronger in
models with large $\sigma_8$, as the clusters are more massive.

In Fig. 4, the plot of $P_2$ versus $P_m$ indeed shows an anticorrelation 
for several
models. Another interesting result is that at fixed $\Omega_0$, $\lambda_0$,
and $\sigma_8$, the magnification probability is insensitive to the value
of $H_0$. This results from the combination of two competing effects.
On one hand, the number of lens planes between the source and the observer
decreases with increasing $H_0$. On the other hand, the total mass in
each plane is proportional to the critical density, and thus increases with 
$H_0$. Hence, models with larger $H_0$ have a smaller number of more
massive lens-planes. 

The histograms of the separation angle (Fig. 5) shows the presence of double
peaks in most cases.  This seems to be more prominent in the models
with larger $\sigma_8$ when the other cosmological parameters are held
fixed. The high density clusters might contain more elliptical galaxies
which have smaller cores than the field galaxies. A more compact core is
known to result in a larger separation angle. Intriguingly, models with smaller
$H_0$ seem to produce longer high separation tails. The presence of
$\lambda_0$ also seems to stretch the histogram to a higher separation tail,
although by a smaller amount than that caused by high $\sigma_8$. 
In addition to this,
it also increases the counts at mid-separation. This might indicate that
structures in $\lambda_0$ models have large sizes which are responsible
for midsize separation angles.

Even at this preliminary stage, the results show the different
tendencies among various cosmological models. With better statistics, we
are hopeful that the differences can be sharpened, and that these
differences will eventually enable us to make meaningful comparisons
 with observational results, thereby 
discriminating the cosmological models.

\section{Cosmological lensing in the direct integration method}
In this section we study the behavior of light ray bundles in 
inhomogeneous model
universes, solving directly the null geodesic equation in the
(cosmological) Newtonian approximation at the stage from an epoch
$t_1$ (with redshift $z_1 = 5$) to the present epoch $t_0$. 
Light rays are deflected by lens objects such as galaxies and
non-galactic clouds. Here, these clouds are mass concentrations
which are unluminous but have masses comparable with a standard
galactic mass. It is not clear at present what average mass and
size and what number density such clouds have, compared with those
 of galaxies. In some previous papers (by Tomita\cite{rf:tom98a}\tocite
{rf:tom98b}), we assumed that non-galactic clouds are dominant and
have a standard galactic mass and lensing strength similar to 
that of galaxies. In this paper we consider another 
case, in which only galaxies are dominant lenses and all non-galactic 
clouds are very weak lenses, and compare the results concerning 
optical deformations in
this case with those in the previous case.  In the future 
observations about the shear deformation at small angles will provide
the information for the structure and lensing strength of non-galactic clouds.

\subsection{Model universes and the ray shooting}

The background model universes have the line-element 
\begin{equation}
  \label{eq:ba1}
ds^2 = -(1+2\varphi/c^2) c^2 dt^2 
+ (1-2\varphi/c^2)a^2(t)(d\mib{x})^2/[1 + K {1 \over 4}(\mib{x})^2]^2,
\end{equation}
where $K$ is the signature of spatial curvature $(\pm 1, \ 0)$. 
The normalized scale factor $S \equiv a(t)/a(t_0)$ satisfies
\begin{equation}
  \label{eq:ba2}
 \Bigl({dS \over d \tau} \Bigr)^2 = {1 \over S} \Bigl[\Omega_0 -
(\Omega_0+\lambda_0-1)S  + \lambda_0 S^3 \Bigr],
\end{equation}
where $\tau \equiv H_0 t$ and $a_0 (\equiv a(t_0))$
is specified by the relation 
$(c H_0^{-1} /a_0)^2 = 1 - \Omega_0 - \lambda_0.$
 The gravitational potential $\varphi$ is described by the Poisson equation
\begin{eqnarray}
  \label{eq:ba3}
a^{-2} \Delta \varphi &=& [1 - {1 \over 4}(\mib{x})^2]^2  \Bigl[{\partial^2 
\varphi \over \partial \mib{x}^2} + {{1 \over 2}x^i \over  [1 + K {1 \over 4}
(\mib{x})^2]^3} {\partial \varphi \over \partial x^i}\Bigr] \cr
&=& 4 \pi G \rho_B [\rho(\mib{x})/\rho_B -1],
\end{eqnarray}
where $\rho_B (=\rho_{B0}/S^3)$ is the background density and 
\begin{equation}
  \label{eq:ba4}
\rho_{B0} = {3 {H_0}^2 \Omega_0 \over 8\pi G} = 2.77 \times 10^{11}
\Omega_0 h^2 M_\odot \ {\rm Mpc}^{-3},
\end{equation}
where $H_0 = 100 h {\rm Mpc^{-1} \ km \ s^{-1}}$. 
In our treatment the inhomogeneities are locally periodic in the
sense that the physical situation at $\mib{x}$ is the same as that at
$\mib{x} + l \mib{n}$, where the components of $\mib{n} 
(=(n^1,n^2, n^3))$ are
integers. In an arbitrary periodic box with coordinate volume $l^3$, 
there are $N$ particles with the same mass $m$. It is assumed that the 
force at an arbitrary point is the sum of forces from $N$ particles in 
the box whose center is the point in question, and that the forces from
outside the box can be neglected. 

In this subsection we consider two flat models (S model and L model) with
$(\Omega_0, \lambda_0) = (1.0, 0)$ and $(0.2, 0.8)$, respectively, and
an open model (O model) with $(0.2, 0)$. The 
present lengths of the boxes are
\begin{equation}
  \label{eq:ba5}
L_0 \equiv a(t_0) l = 32.5 h^{-1}, \ 50 h^{-1}, \ 50 h^{-1}{\rm Mpc}
\end{equation}
for S, L and O models, respectively. The particle number is $32^3$ 
in all models, and thus
\begin{equation}
  \label{eq:ba6}
m (= \rho_{B0} {L_0}^3/N) = 2.90, \ 2.11, \ 2.11 \times 10^{11} h^{-1} M_\odot,
\end{equation}
respectively.

The distributions of particles in these models were derived using numerical
$N$-body simulations.
The particle size (giving the lens strength) is represented by the 
softening radius $r_s~(= a(t) x_s)$, which is constant. For the
particle size we consider two lens models for comparison:

\noindent {\bf Lens model 1}. All particles in the low-density models 
($\Omega_0 = 0.2$) have $r_s = 20 h^{-1}$kpc,  $20 \%$ of the particles in
the flat model $(1.0, 0)$ have $r_s = 20 h^{-1}$kpc, and the
remaining particles have  $r_s = 500 h^{-1}$kpc. 
Thus practically, particles with $\Omega_c = 0.2$ (which we call {\it compact 
lens objects} in previous papers\cite{rf:tom98a}\tocite{rf:tom98b})
 play the role of lens objects. Their number density is
much larger than the galactic density $\Omega_g \sim 0.02$. 

\noindent {\bf Lens model 2}. 
$10 \%$ of the particles in the low-density model ($\Omega_0 = 0.2$)
and $2 \%$ of the particles in the flat model $(1.0, 0)$ have 
$r_s = 20 h^{-1}$kpc, while the remaining particles have 
 $r_s = 500 h^{-1}$kpc. Thus only galaxies corresponding to
$\Omega_g = 0.02$ play significant roles as lens objects, and the remaining
particles are regarded as diffuse clouds.

 The lens strength of realistic non-galactic clouds is
intermediate between these two lens models. The question of which 
model is better may be answered by observational studies involving
 lens phenomena.

The time evolution of the distribution of particles was derived by
performing the $N$-body simulation in the tree-code provided by
Suto.\cite{rf:su}
The initial particle distributions were derived using Bertschinger's
software {\it COSMICS}\cite{rf:bert}
 under the condition that their perturbations are given
as random fields with the spectrum of cold dark matter, their power
$n$ is 1, and their normalization is specified as the dispersion 
$\sigma_8 \ = \ 0.94$ with the Hubble constant $h = 0.5$ for $(1.0,
0)$ and $h = 0.7$ for other models with $\Omega_0 = 0.2$.

Now let us consider light propagation described by the null geodesic
equation with the null condition. Here we use $T \equiv 
{1 \over 2}\ln
[a(t)/a(t_1)]$ as a time variable and $T_0 \equiv (T)_{t = t_0}$. Then
we have $dS = 2 \exp [2(T- T_0)] dT$, so that 
\begin{equation}
  \label{eq:ba7}
c dt = R\ c_R \ [\Omega_0 +(1 -\Omega_0 -\lambda_0) S + \lambda_0 
S^3]^{-1/2} {e}^{3 T} dT,
\end{equation}
where
\begin{equation}
  \label{eq:ba8}
R \equiv L_0/[(1+z_1) N^{1/3}]
\end{equation}
and
\begin{equation}
  \label{eq:ba9}
c_R \equiv 2(c/H_0)/[R(1+z_1)^{3/2}].
\end{equation}
The line-element is 
\begin{eqnarray}
  \label{eq:ba10}
ds^2/R^2 = -{c_R}^2  [\Omega_0 &+&(1 -\Omega_0 -\lambda_0) S + 
\lambda_0 S^3]^{-1} {e}^{6T} (1+\alpha \phi) dT^2 \cr 
&+& (1-\alpha \phi) {e}^{4T} d\mib{y}^2 /F(\mib{y})^2,
\end{eqnarray}
where \ $y^0 \equiv T, \ y^i \equiv a(t_1) x^i/R, \ \varphi \equiv
 (G m/ R) \phi$, \ $R_0 \equiv R a_0/a_1 = (1+z_1) R$, 
\begin{equation}
  \label{eq:ba11}
\alpha \equiv {2Gm \over c^2 R} = {3 \over \pi} {\Omega_0 \over (c_R)^2}
\end{equation}
and
\begin{equation}
  \label{eq:ba11a}
F \equiv 1 -{1 \over 4}(R_0H_0/c)^2 (1 -\Omega_0 -\lambda_0) 
(\mib{y})^2.
\end{equation}
The equations to be solved for light rays are 
\begin{equation}
  \label{eq:ba12}
{dy^i \over dT} = c_R  {e}^T \tilde{K}^i,
\end{equation}
\begin{eqnarray}
  \label{eq:ba13}
{d\tilde{K}^i \over dT} = &-&[3 \lambda_0 e^{4(T-T_0)}+(1 -\Omega_0 
-\lambda_0)]  e^{2(T-T_0)} \tilde{K}^i /G(T)
  +\alpha {\partial \phi 
\over \partial T} \tilde{K}^i \cr
&-&\gamma {c_R}^{-1}  {e}^T \Bigl[{\partial \phi /
\partial y^i}/G(T) - 2 {\partial \phi \over
\partial y^j} \tilde{K}^j \tilde{K}^i\Bigl] \cr
&+& (R_0H_0/c)^2 (1 -\Omega_0 -\lambda_0) F^{-1} c_R e^T   \cr
&&\times \Bigl[-y^j \tilde{K}^j \tilde{K}^i 
 + {1 \over 2}(1 +2\alpha \phi)/G(T) \Bigl] ,
\end{eqnarray}
where $\tilde{K}^i \equiv {c_R}^{-1} e^{-T} {dy^i / dT}, \ 
 \gamma \equiv \alpha (c_R)^2$ \ and
\begin{equation}
  \label{eq:ba13a}
G(T) = \Omega_0 +(1 -\Omega_0 
-\lambda_0) e^{2(T-T_0)}+ \lambda_0  e^{6(T-T_0)}.
\end{equation}
The null condition is 
\begin{equation}
  \label{eq:ba14}
\sum_i (\tilde{K}^i)^2 = 1 + 2\alpha \phi.
\end{equation}

While in flat models the solution is straightforward, some care must
be taken in curved models. 
The potential $\phi$ is given as a solution of the Poisson equation.
Because the ratio of the second term to the first term on the
right-hand side of Eq. (\ref{eq:ba3}) is $(R_0H_0/c)^2 (\mib{y})^2
[\Delta y/\vert \mib{y}\vert] \ll 1$ for $z \leq z_1$, the Poisson
equation in a box can be approximately expressed as
\begin{equation}
  \label{eq:ba15}
F(\mib{y}_c)^2 \ {\partial^2 \phi \over \partial \mib{y}^2} =
4 \pi G \rho_B [\rho(\mib{x})/\rho_B -1],
\end{equation}
where we have used $F(\mib{y}) \simeq 1$ for $y \sim$ [the box size].
$F(\mib{y})$ can be approximately replaced by the central value 
$F(\mib{y}_c)$ for $y \gg $ [the box size], where the index $c$ denotes the central value in the box. 
For point sources with $ \rho  = m \sum_n \delta(a R (\mib{y} 
-{\mib{y}_n}))$ \ ($n$ is the particle number), we have 
\ $\phi = \phi_1 + \phi_2$, \ where   
\begin{equation}
  \label{eq:ba16}
\phi_1 = - F(\mib{y}_c) e^{-2T} \sum_n {1 \over \vert \mib{y} 
-\mib{y}_n\vert},
\end{equation}
and $\phi_2$ represents the contribution from the homogeneous background
density. Here let us use for $\mib{y}$ another vectors $\bar{\mib{y}}$
expressing the space in a locally flat manner, where the two coordinates
are related as 
\begin{equation}
  \label{eq:ba17}
\bar{\mib{y}} = \int^{\mib{y}}_{\bf 0} d\mib{y}/F(\mib{y}),
\end{equation}
and the lengths between two points in boxes in two coordinates are 
approximately related as
\begin{equation}
  \label{eq:ba18}
\Delta \bar{\mib{y}} = \Delta \mib{y}/F(\mib{y}_c).
\end{equation}
Then $\phi_1$ can be expressed in terms of $\bar{\mib{y}}$ in the usual
manner as
\begin{equation}
  \label{eq:ba19}
\phi_1 = - e^{-2T} \sum_n {1 \over \vert \bar{\mib{y}} 
-\bar{\mib{y}}_n\vert}.
\end{equation}
Corresponding forces are expressed as $f_i \equiv \partial
\phi_1/ \partial y^i = [\partial \phi_1/ \partial {\bar y}^i]/F(\mib{y}_c)$.
It should be noted that the contribution of $\partial \phi/ \partial T$ is 
negligibly small, compared with that of $f_i$.

In flat models the universe is everywhere covered with periodic boxes
continuously connected as in Fig. 6. In open models it cannot be covered
in such a way, but we can consider only a set of local periodic boxes
connected along each light ray, as in Fig. 7. In these boxes we can 
describe the evolution in the distribution of particles in terms of 
local flat coordinates $\bar{\mib{y}}$, because the size of boxes is 
much smaller than the curvature radius. 

For the integration of the above null-geodesic equations, we calculate 
the potential at a finite number of points on the ray which are 
given at each time step ($\Delta T$). Then particles near one of the 
points on the ray have a stronger
influence upon the potential than particles far from any points. 
To avoid this unbalance in the calculation of the potential, we take
an average of the potential $\phi_1$ by integrating it analytically 
over the interval between one of the points and the next point. 
The expression for an averaged potential $\bar \phi$ was given in a 
previous paper. Moreover, to take into account the finite particle
size as galaxies or non-galactic clouds,
we modify the above potential for point sources using the 
softening radii $y_s = a(t_1) x_s/R$. The modified potential is
produced by replacing $(y-y_n)^2$ by $(y-y_s)^2 + (y_s)^2$ in the  
potential for point sources.  

The initial values of $\tilde{K}^i$ are given so as to satisfy
Eq. (\ref{eq:ba14}). The integration of Eqs. (\ref{eq:ba12}) and
(\ref{eq:ba13}) with the modified potential was performed using the Adams 
method. As the time step we used $\Delta
T = [\ln 6/2]/N_s$ with $N_s = 3000$ (in most cases) $ - 10000$. 

Here we consider the comoving volume of the region through which a ray 
bundle with solid angle $\theta^2$ pass at the interval $0 \leq z \leq 
z_i$ in the flat model $(1.0, 0)$. This volume is equal to the volume 
of the cone $V_R = {1 \over 3}
\theta^2 \ [2 c H_0 (1 - 1/\sqrt{1+z_i})]^3$. Its ratio
to the volume of the periodic box $V_B \ [= (32 h^{-1}{\rm Mpc})^3]$ is
$1.2 \times 10^{-3}$ for $z_i = 5$ and $\theta = 10$arcsec. This small 
ratio implies that the influence of the periodicity of the box on the
the statistics of ray bundles is negligibly small, as long as the ray
bundles are not directed in some special directions with respect to the box. 

\begin{figure}[htb]
 \parbox{\halftext}{
\epsfxsize=5.5cm
\epsfbox{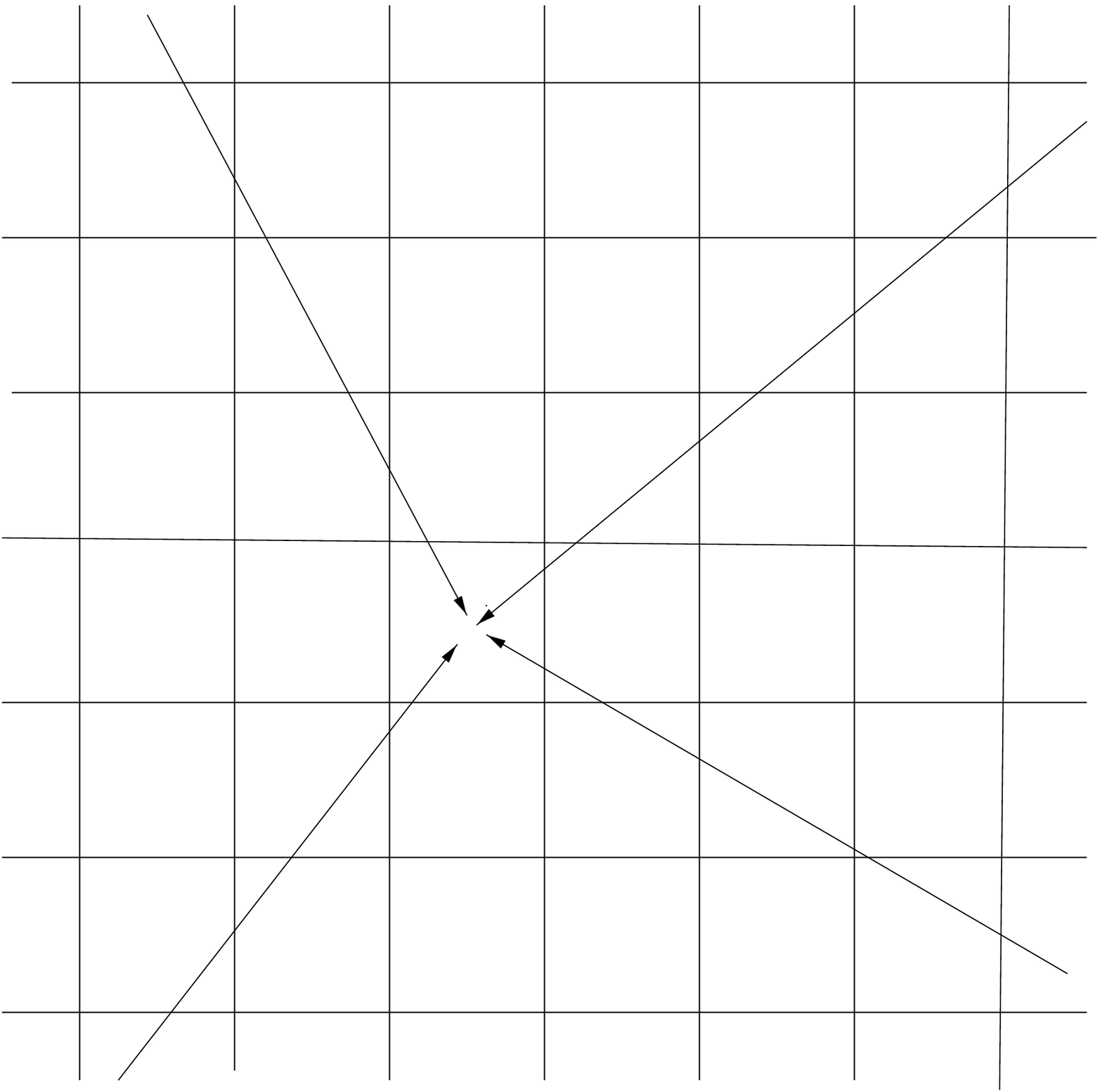}
\caption{Light rays and periodic boxes in a flat space.}}
\hspace{1mm}
 \parbox{\halftext}{
\epsfxsize=4.3cm
\epsfbox{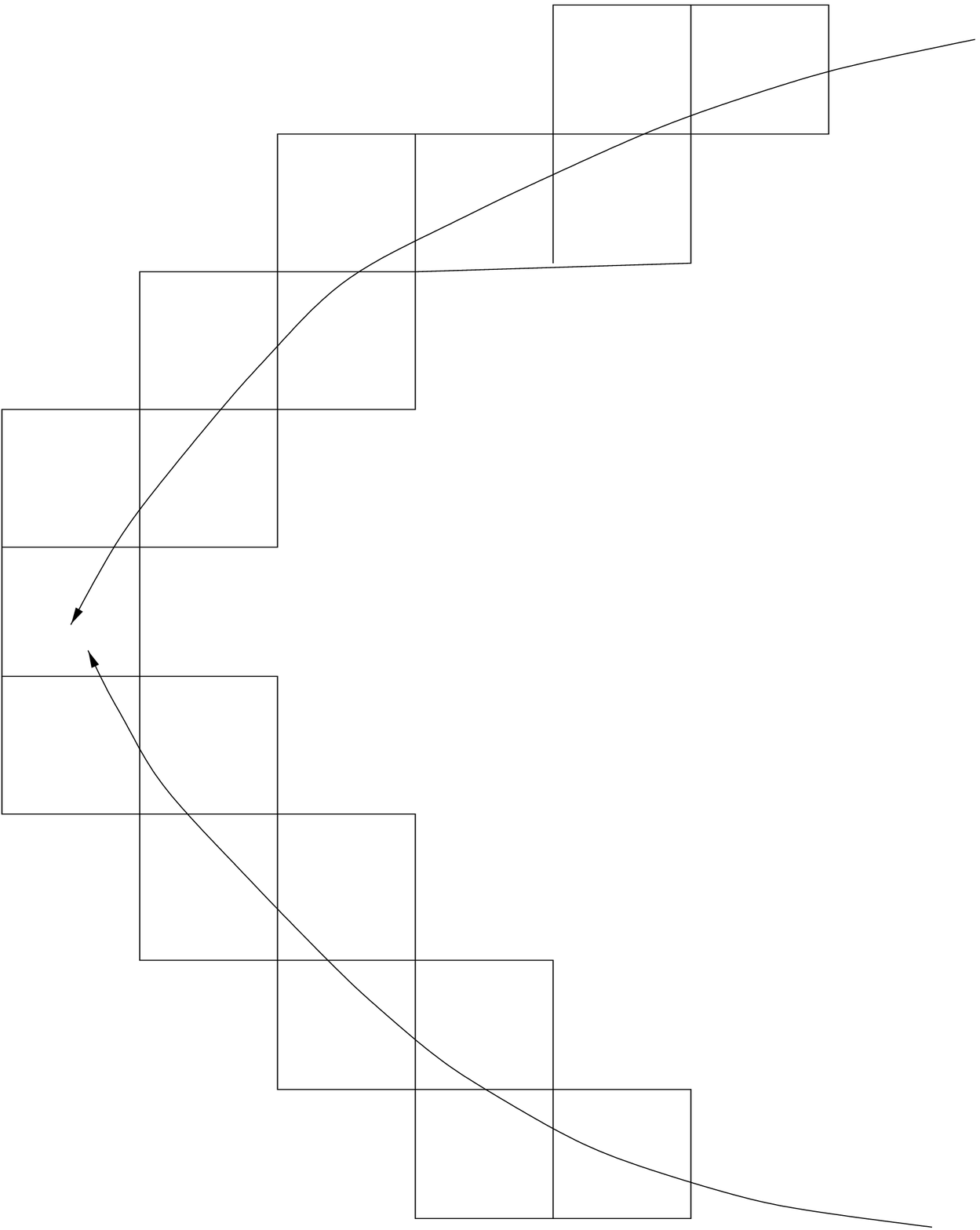}
\caption{Light rays and periodic boxes in an open space.}}
\end{figure}

\subsection{Statistical behavior of optical scalars}

We treat the deformation of ray bundles over the interval from $z = 0$ 
to $z = 5$ measured by an observer in a periodic box. Here we consider 
the ray bundles reaching the observer (or emitted backwards) in a
regular form such that 
the rays are put in the same separation angle $\theta$, and
calculate the relative change in the angular positions of the rays
which increases
with redshift in the past direction. From this change we find the
behavior of optical scalars.

Basic ray bundles consist of  5 $\times$ 5  rays that are put in a square
form with the same
separation angle $\theta = 2 - 360$ arcsec. Many bundles coming 
from all directions in
the sky are considered. Here we take 200 bundles coming from randomly
chosen directions for each separation angle. In order to express
relative angular positions of rays, we use two orthogonal vectors, 
$e^i_{(1)}$ and $e^i_{(2)}$, in the plane perpendicular to the first 
(fiducial) background ray vector $(\tilde{K}^i)_B$.  
Then the angular coordinates $[X(m,n), Y(m,n)]$ of 25 rays relative to 
the first (fiducial) ray with $(m,n) = (1,1)$ at any epoch are defined by
\begin{eqnarray}
  \label{eq:st2}
X(m,n) &=& \sum_i [y^i(m,n) -  y^i(1,1)]\ e^i_{(1)}/y_B (1,1) +X(1,1),\cr
Y(m,n) &=& \sum_i [y^i(m,n) -  y^i(1,1)]\ e^i_{(2)}/y_B (1,1) +Y(1,1),
\end{eqnarray}
where $y^i = y^i_B + \delta y^i$ and $y_B = [\sum_i (y^i_B)^2]$.
Since all angular intervals of rays at the observer's point are the same ($=
\theta$), the differentiation of angular coordinates of the
rays at any epoch with respect to those at the observer's points is given
by the differences
\begin{eqnarray}
  \label{eq:st3}
A_{11}(m,n) &=& [X(m+1,n) - X(m,n)]/\theta, \cr
A_{12}(m,n) &=& [X(m,n+1) - X(m,n)]/\theta, \cr
A_{21}(m,n) &=& [Y(m+1,n) - Y(m,n)]/\theta, \cr
A_{22}(m,n) &=& [Y(m,n+1) - Y(m,n)]/\theta, 
\end{eqnarray}
where $m$ and $n$ run from 1 to 5. From the matrix $A_{ij} (m,n)$ we
derive the optical scalars in the standard manner,\cite{SCH92}
as the convergence ($\kappa (m,n)$),
the shear ($\gamma_i (m,n), \ i=1,2$), and the amplification ($\mu
(m,n)$) defined by
\begin{eqnarray}
  \label{eq:st4}
 \kappa (m,n) &=& 1 - {\rm tr}(m,n)/2, \quad \gamma_1 (m,n) = [A_{22}(m,n) - 
A_{11}(m,n)]/2, \cr
\gamma_2 (m,n) &=& -[A_{12}(m,n) + A_{21}(m,n)]/2, \cr
\gamma^2 &\equiv& (\gamma_1)^2 + (\gamma_2)^2 = [{\rm tr}(m,n)]^2 - 
\det(A_{ij}(m,n)), \cr
\mu (m,n) &=& 1/\det(A_{ij}(m,n)),
\end{eqnarray}
where the trace  is ${\rm tr}(m,n) = A_{11}(m,n) + A_{22}(m,n)$.
The average optical quantities in each bundle are defined as the
averages of optical quantities for all rays in the bundle as 
\begin{equation}
  \label{eq:st5}
\bar{\kappa} = \Bigl[\sum_m \sum_n \kappa (m,n)\Bigr]/4^2, \quad
\bar{\kappa^2} = \Bigl[\sum_m \sum_n (\kappa (m,n))^2\Big]/4^2,
\end{equation}
and so on. In this averaging process the contributions from smaller 
scales can be cancelled and smoothed-out. The above optical scalars 
at the separation angle $\theta$ are accordingly derived in the 
coarse-graining on this smoothing scale. 

For the present statistical analysis we excluded caustic cases,
considering only the case of {\it weak lensing} in the sense of
{\it no caustics}.
The averaging for all non-caustic ray bundles is denoted using
$\langle \rangle$
as $\langle\kappa^2\rangle, \ \langle\gamma^2\rangle$  and  
$\langle(\mu -1)^2\rangle$. Because
$\kappa(m,n), \ \gamma_i (m,n)$ and $\mu (m,n) -1$ take positive and
negative values with almost equal frequency, $\langle\kappa\rangle,
 \ \langle\gamma_i\rangle$ 
and $\langle\mu\rangle -1$  are small. 

In Figs. 8 and 9, we show the behavior of $\langle\gamma^2\rangle$ for various 
separation angles $\theta = 2$ arcsec  $- \ 360$ arcsec (= $6$ arcmin) 
at epochs $z = 1.0$ and $2.0$, respectively. Similarly in Figs. 10 and
11, we show the behavior of $\langle(1 - \mu)^2\rangle$ at epochs 
$z = 1.0$ and 
$2.0$, respectively. The calculations were
performed at $\theta = 2, 5, 10, 30, 60,$ and $360$ arcsec.
The behavior of  $\langle\kappa^2\rangle$ is similar to that of $\langle
\gamma^2\rangle$, 
as was shown in previous papers. In all figures, the values in the two 
lens models are shown using solid and dotted lines.  
It is found in the low-density models that the ratios of $\langle
\kappa^2\rangle^{1/2}$, 
$\langle\gamma^2\rangle^{1/2}$ and $\langle(1 - \mu)^2
\rangle^{1/2}$ in the lens model 1
to those in the lens model 2 are about 4 at the separation angle $\theta
= 2$ arcsec. The ratios decrease with the increase of $\theta$, and are 
$< 2$ at $\theta = 360$ arcsec. That is, the remarkable difference between 
the two lens
models appears at the small angles $\theta \simeq 2$ arcsec.
The values of optical scalars for $\theta \sim 360$ arcsec are roughly
consistent with the results of Bernardeau et al. (cf. their Figs. 
3 and 4)\cite{rf:bern97} and Nakamura\cite{rf:nakam97} in their treatments.
  
At all separation angles and for both lens models, $\langle\kappa^2\rangle, 
\langle\gamma^2\rangle$ and $\langle(1 - \mu)^2\rangle$ in the open 
model ($\Omega_0 = 0.2$) 
are larger than those in the flat model (L). In model S they
are largest at all angles for the lens model 2 and at large angles 
($\theta \sim 360$ arcsec) in the lens model 1, but they are smallest
at small angles ($\theta < 30$ arcsec) because the role of
non-galactic clouds on small-scale is small. 

In Tables I, II and III, we show for models S, L and O the 
numerical values of $\langle\kappa\rangle,\ 
\langle\kappa^2\rangle^{1/2}, \ \langle\mu\rangle, \ \langle(\mu -1)^2
\rangle^{1/2}, \  \langle\gamma_1\rangle$ and 
$\langle\gamma^2\rangle^{1/2}$ \ at epochs $z = 1,2, .., 5$ for $\theta = 2$ 
arcsec. 

The influence of lensing on source magnitudes is 
\begin{equation}
  \label{eq:st6}
\Delta m = {5 \over 2} \log [1 + \langle(1-\mu)^2\rangle^{1/2}] \simeq 1.09 
\langle(1-\mu)^2\rangle^{1/2}.
\end{equation}
If $\theta = 2$ arcsec, the value of $\Delta m$ is $(0.058, 0.034), 
(0.077, 0.019)$ and $(0.133, 0.022)$ for the 
lens model $(1, 2)$, respectively, at $z = 1$ in models S, L and O. 
If the lens model 1 is more realistic, therefore, this lens correction may 
introduce some sensitive modification to the standard selection of cosmological
models in the $[m, z]$ relation. Here it should be noted that the
lensing effect at small angles, which is caused by small-scale
nonlinear inhomogeneities, is important for the magnitude correction
of small high-redshift objects, such as SNe Ia, in cosmological 
observations.\cite{rf:perl}\tocite{rf:riess}

\begin{figure}
\epsfxsize=7.5cm
\centerline{\epsfbox{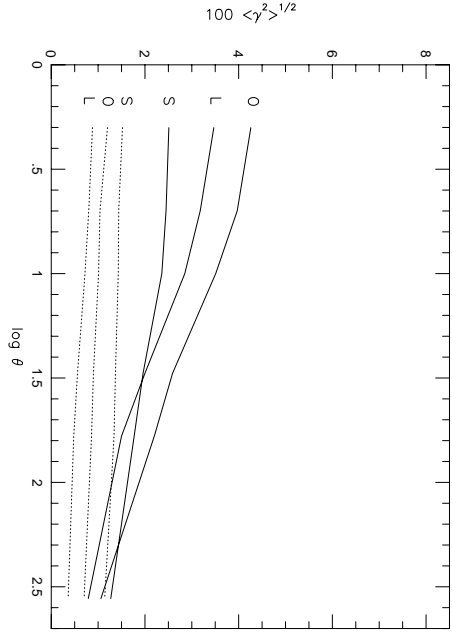}}
\caption{The angular dependence of $\langle\gamma^2\rangle^{1/2}$. 
Solid and dotted lines represent behavior for lens models 1 and 2,
respectively, at $z = 1$.
The separation angle $\theta$ is in units of arcsec.}
\label{fig:8}
\end{figure}

\begin{figure}
\epsfxsize=7.5cm
\centerline{\epsfbox{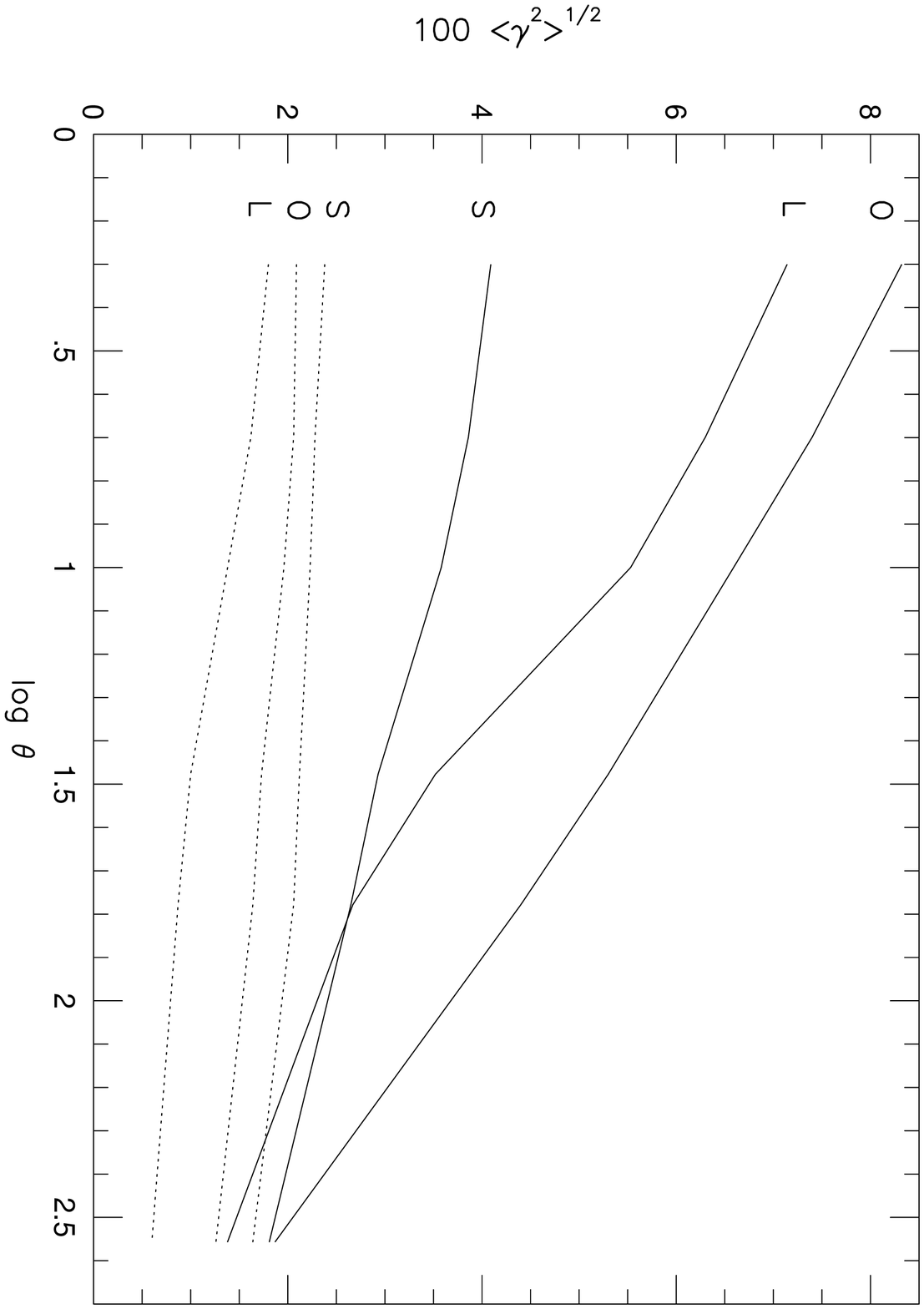}}
\caption{The angular dependence of $\langle\gamma^2\rangle^{1/2}$.
Solid and dotted lines denote behavior for lens models 1 and 2,
respectively, at $z = 2$. 
The separation angle $\theta$ is in the unit of arcsec.}
\label{fig:9}
\end{figure}

\begin{figure}
\epsfxsize=7.5cm
\centerline{\epsfbox{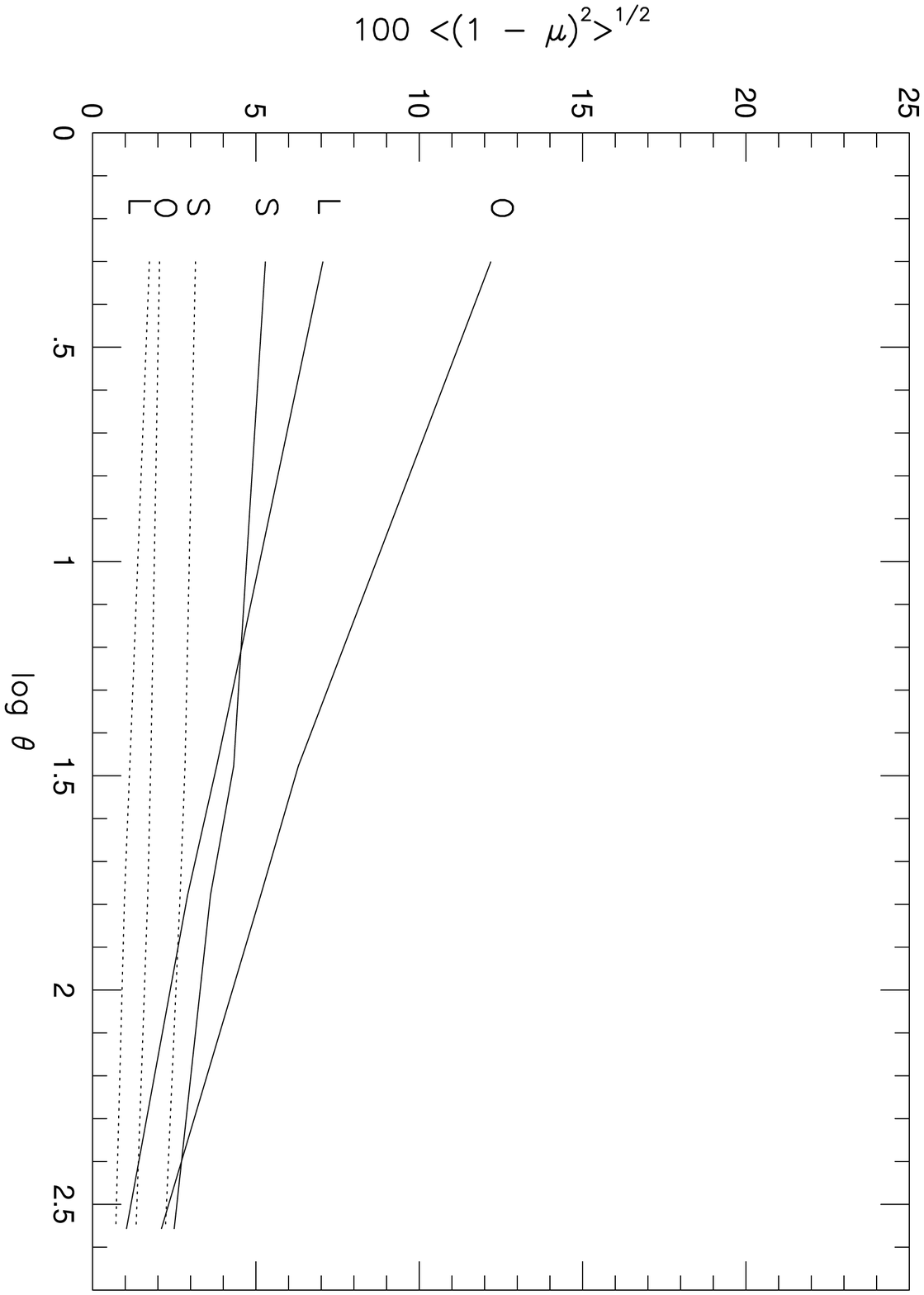}}
\caption{The angular dependence of $\langle(1 - \mu)^2\rangle^{1/2}$. 
Solid and dotted lines represent behavior for lens models 1 and 2,
respectively, at $z = 1$.
The separation angle $\theta$ is in units of arcsec.}
\label{fig:10}
\end{figure}

\begin{figure}
\epsfxsize=7.5cm
\centerline{\epsfbox{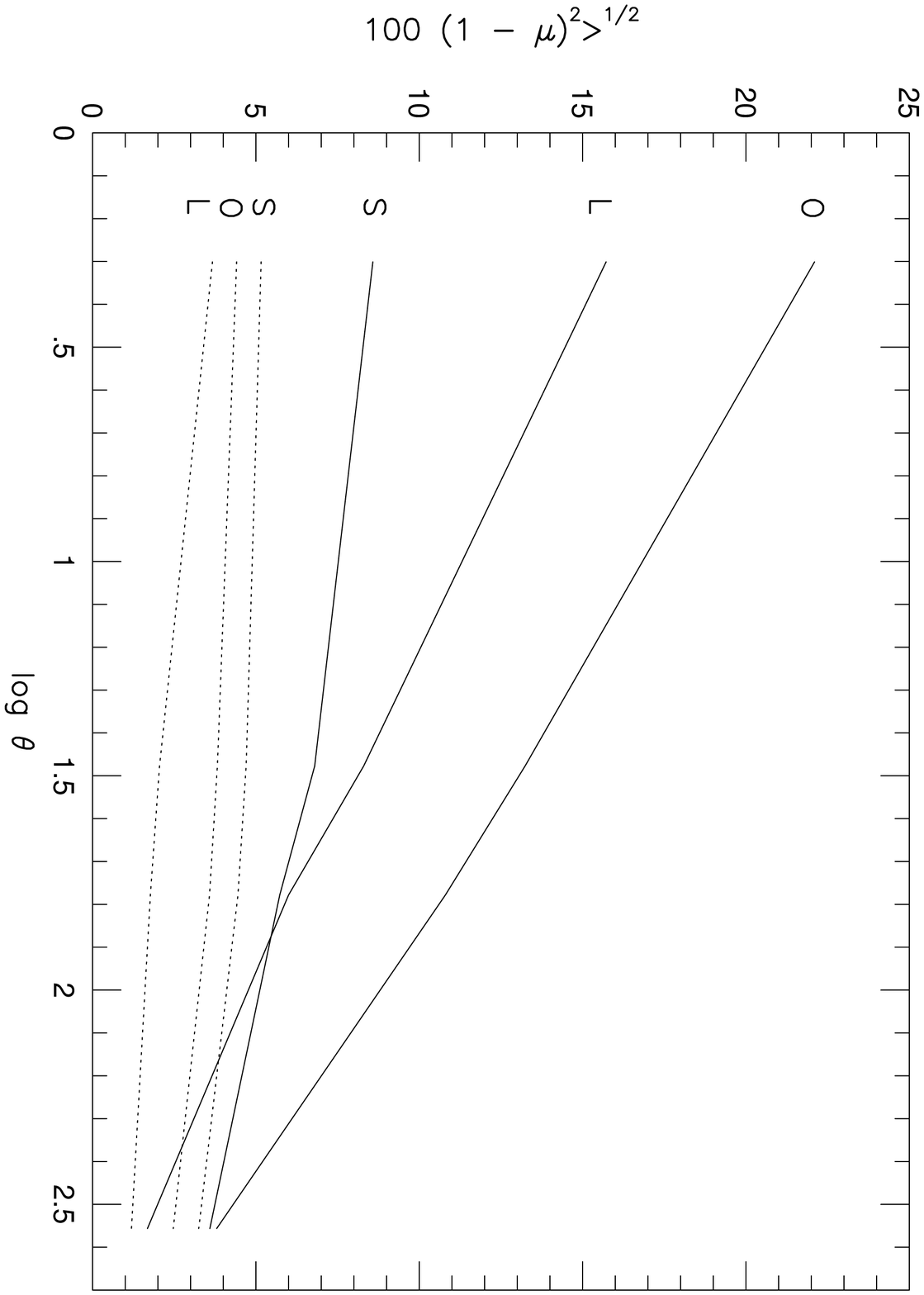}}
\caption{The angular dependence of $\langle(1 - \mu)^2\rangle^{1/2}$.
Solid and dotted lines represent behavior for lens models 1 and 2,
respectively, at $z = 2$. 
The separation angle $\theta$ is in the unit of arcsec.}
\label{fig:11}
\end{figure}

\begin{table}
\caption{Optical quantities in model S at $\theta = 2$ arcsec.} 
\label{table:1}
\begin{center}
\begin{tabular}{ccccccrc} \hline \hline
$lens$&$z$&$\langle\kappa\rangle$ & $\langle\kappa^2\rangle^{1/2}$  & 
$\langle\mu\rangle$ & $\langle(1-\mu)^2\rangle^{1/2}$
 & $\langle\gamma_1\rangle$ & $\langle\gamma^2\rangle^{1/2}$
 \\ \hline 
&1&   $ 0.0005$ &  $0.0245$ & 1.0036 & 0.0529 & $-0.0032$ & 0.0251\\
&2&   $ 0.0004$ &  $0.0398$ & 1.0075 & 0.0858 & $-0.0061$ & 0.0409\\
1&3&  $ 0.0021$ &  $0.0495$ & 1.0149 & 0.1084 & $-0.0080$ & 0.0514\\
&4&   $ 0.0034$ &  $0.0563$ & 1.0208 & 0.1251 & $-0.0096$ & 0.0593\\
&5&   $ 0.0042$ &  $0.0617$ & 1.0253 & 0.1390 & $-0.0109$ & 0.0653\\ \hline
&1&   $ 0.0005$ &  $0.0155$ & 1.0020 & 0.0315 & $-0.0026$ & 0.0152\\
&2&   $ 0.0020$ &  $0.0251$ & 1.0065 & 0.0516 & $-0.0049$ & 0.0238\\
2&3&  $ 0.0030$ &  $0.0305$ & 1.0096 & 0.0636 & $-0.0053$ & 0.0283\\
&4&   $ 0.0037$ &  $0.0343$ & 1.0120 & 0.0721 & $-0.0055$ & 0.0314\\
&5&   $ 0.0041$ &  $0.0370$ & 1.0136 & 0.0784 & $-0.0055$ & 0.0339\\ \hline
\end{tabular}
\end{center}
\end{table}

\begin{table}
\caption{Optical quantities in model L at $\theta = 2$ arcsec. } 
\label{table:2}
\begin{center}
\begin{tabular}{ccccccrc} \hline \hline
$lens$&$z$&$\langle\kappa\rangle$ & $\langle\kappa^2\rangle^{1/2}$  & 
$\langle\mu\rangle$ & $\langle(1-\mu)^2\rangle^{1/2}$
 & $\langle\gamma_1\rangle$ & $\langle\gamma^2\rangle^{1/2}$
 \\ \hline 
&1&   $-0.0011$ &  $0.0324$ &  1.0025 &  0.0705 &  0.0008 &  0.0347\\
&2&   $-0.0036$ &  $0.0675$ &  1.0134 &  0.1572 &  0.0018 &  0.0714\\
1&3&   $-0.0045$ & $0.0974$ &  1.0345 &  0.2407 &  0.0020 &  0.1002\\
&4&   $-0.0065$ &  $0.1216$ &  1.0560 &  0.3209 &  0.0023 &  0.1214\\
&5&   $-0.0070$ &  $0.1405$ &  1.0805 &  0.4000 &  0.0030 &  0.1372\\ \hline
&1&   $ 0.0001$ &  $0.0084$ &  1.0004 &  0.0174 & $-0.0004$ &  0.0088\\
&2&   $ 0.0004$ &  $0.0175$ &  1.0021 &  0.0367 & $-0.0018$ &  0.0180\\
2&3&  $ 0.0001$ &  $0.0239$ &  1.0025 &  0.0505 & $-0.0020$ &  0.0245\\
&4&   $ 0.0001$ &  $0.0291$ &  1.0038 &  0.0618 & $-0.0019$ &  0.0296\\
&5&   $ 0.0005$ &  $0.0333$ &  1.0055 &  0.0710 & $-0.0017$ &  0.0334\\ \hline
\end{tabular}
\end{center}
\end{table}

\begin{table}
\caption{Optical quantities in model O at $\theta = 2$ arcsec. } 
\label{table:3}
\begin{center}
\begin{tabular}{ccccccrc} \hline \hline
$lens$&$z$&$\langle\kappa\rangle$ & $\langle\kappa^2\rangle^{1/2}$
  & $\langle\mu\rangle$ & $\langle(1-\mu)^2\rangle^{1/2}$
 & $\langle\gamma_1\rangle$ & $\langle\gamma^2\rangle^{1/2}$
 \\ \hline 
&1&   $-0.0002$ &  $0.0425$ &  1.0221 &  0.1219 & $-0.0028$ &  0.0426\\
&2&   $-0.0054$ &  $0.0796$ &  0.9961 &  0.2210 & $-0.0041$ &  0.0832\\
1&3&  $-0.0069$ &  $0.1157$ &  1.0368 &  0.6234 & $ 0.0057$ &  0.1225\\
&4&   $-0.0081$ &  $0.1519$ &  1.0790 &  0.5638 & $ 0.0078$ &  0.1602\\
&5&   $-0.0064$ &  $0.1872$ &  1.1627 &  1.1880 & $ 0.0120$ &  0.1980\\ \hline
&1&   $-0.0003$ &  $0.0099$ &  0.9997 &  0.0205 & $-0.0005$ &  0.0120\\
&2&   $-0.0012$ &  $0.0209$ &  0.9993 &  0.0441 & $-0.0015$ &  0.0209\\
2&3&  $ 0.0000$ &  $0.0314$ &  1.0041 &  0.0682 & $-0.0035$ &  0.0284\\
&4&   $ 0.0010$ &  $0.0410$ &  1.0089 &  0.0915 & $-0.0047$ &  0.0370\\
&5&   $ 0.0021$ &  $0.0496$ &  1.0144 &  0.1140 & $-0.0059$ &  0.0451\\ \hline
\end{tabular}
\end{center}
\end{table}

For the statistical analysis we used 200 ray bundles reaching an
observer in a single inhomogeneous model universe. This number of ray
bundles may be too small to cover the influences from complicated
inhomogeneities in all directions. To obtain more robust statistical 
results it may be necessary to use more ray bundles and more model
universes produced with random numbers.

Finally, we touch upon recent observations of cosmological shear due to weak
lensing and their relation to our results.  Fort et al.\cite{rf:fort}
 attempted  
measurements of a coherent shear from foreground mass condensations in 
the fields of several luminous radio sources.  Schneider et
al. \cite{rf:sch}
determined the shear in the field ($2$ min $\times 2$ min)
containing a radio source PKS1508-05 with $z = 1.2$, and their result
is that the shear is about 0.03 for an angular scale of 1 min.
This value may be consistent with the low-density models in the lens
model 1, but more observational data are necessary to deduce any
conclusions. 
In order to clarify the lensing strength of
non-galactic clouds it is important to have observational values
of the shear at small angles $\theta \approx 2$ arcsec.

\newcommand{\bm}[1]{\mbox{\boldmath $#1$}}

\section{Analytic evaluation of cosmic shear through the power spectrum
approach \protect\footnote{In this section we use units in which $c=H_0=1$
and follow Misner et al.\protect\cite{MIS73} as regards the sign convention
of curvature tensors.} }\label{sec:ttn}

In this section we analytically estimate the shear and the image
amplification caused by large scale density inhomogeneity in the
universe.{\cite{rf:gunnb}~\cite{rf:bl91}\tocite{rf:nakam97}
emanating from us to the past, and let $\chi$ be the affine parameter
along the central ray in the bundle and $k^a = (d/d\chi)^a$ be its
tangent vector. Deformation in the cross section of the bundle is
described by optical scalars, the expansion $\theta = \frac12
\nabla_{\!\!a} k^a$ and shear $\sigma = \frac12 \epsilon^a \epsilon^b
\nabla_{\!\!a} k_b$, where $\epsilon^a = e_1^a + i e_2^a$ is the complex
dyad basis on the cross section.  The rotation $\omega = \frac12 {\rm
Im} ( \epsilon^{*a} \epsilon^b ) \nabla_{\!\!a} k_b$ identically
vanishes for a bundle which emanates from a single point, so we set
$\omega=0$ in the following. These evolve along the ray according to the
equations \cite{SCH92,SAS93}
\begin{equation}\label{n1}
  \dot\theta + \theta^2 + |\sigma|^2 = -{\cal R} \,,\quad \dot\sigma +
  2\theta\sigma = -{\cal F} \,,
\end{equation}
where the overdot denotes $d/d\chi$, and the sources of deformation are
${\cal R} = \frac12 R_{ab} k^a k^b$ and ${\cal F} = \frac12 C_{acbd}
\epsilon^a \epsilon^b k^c k^d$. The observable effect of this
deformation is the evolution of a deviation vector $\xi^a$ connecting
two nearby rays on the cross section:
\begin{equation}
  \dot \xi = \theta\xi + \sigma \xi^*.
\end{equation}
where $\xi=\epsilon_a\xi^a$. This deviation vector is written as a
linear mapping from the initial value of $\dot\xi$ at $\chi=0$:
\begin{equation}\label{n3}
  \xi = \Lambda^* \dot\xi_0 + \Gamma \dot \xi_0^*.
\end{equation}
Here $\dot\xi_0$ represents the angular separation between the 
nearby rays in the observer's sky, and ${\rm Re}\Lambda$, ${\rm
Im}\Lambda$ and $\Gamma$ represent respectively the degrees of focusing,
rotation and shear in the shape of the cross section as seen from the
observer. From equation (\ref{n1}) it follows that
\begin{equation}\label{n4}
  \dot\Lambda  = \theta\Lambda + \sigma^* \Gamma \,,\quad 
  \ddot\Lambda = -({\cal R}\Lambda + {\cal F}^*\Gamma ), 
\end{equation}
\begin{equation}\label{n5}
  \dot\Gamma  = \sigma \Lambda + \theta\Gamma \,,\quad
  \ddot\Gamma = -({\cal F}\Lambda + {\cal R} \Gamma).
\end{equation}
The angular diameter distance $D$ is defined to be (cross sectional
area/solid angle in the sky)$^{1/2}$ so $D = (|\Lambda|^2 -
|\Gamma|^2)^{1/2}$ and satisfies
\begin{equation}
  \dot D = \theta D \,,\quad \ddot D = -({\cal R} + |\sigma|^2) D.
\end{equation}
Initial conditions for these differential equations at $\chi=0$ are $0=
\Lambda= \Gamma= \dot\Gamma= D= \sigma$ and $\dot\Lambda= \dot D =1$.

The following form of the metric is used to describe the globally
homogeneous but locally inhomogenous universe
\begin{equation}\label{n7}
  d\hat s^2 = a^2 ds^2 = a^2(\eta)\,[ -(1+2\Phi) d\eta^2 + (1-2\Phi)
  \gamma_{ij} dx^i dx^j ].
\end{equation}
Here $a$ is the scale factor and $\gamma_{ij}$ is the 3-metric on the
constant time hypersurface of curvature $K=\Omega_0+\lambda_0-1$.  The
gravitational potential $\Phi$ ($\ll 1$) incorporates the local
inhomogeneity generated by the density contrast $\delta =
(\rho-\bar\rho)/\bar\rho$ and satisfies the Poisson equation
\begin{equation}\label{n8}
  \nabla^2 \Phi = \frac32\Omega_0 \frac\delta a,
\end{equation}
where $\nabla^2$ is the Laplacian associated with $\gamma_{ij}$ (we only
consider a sub-horizon scale inhomogeneity whose wavelength is much
smaller than $|K|^{-1/2}$). Since the null geodesics are unaffected by
the conformal transformation, we work in the scaled metric $ds^2$ in
what follows. Accordingly the ``distances'' $\hat\Lambda$ and
$\hat\Gamma$ (Eq.(\ref{n3})) in the metric $d\hat s^2$ are related to
those in $ds^2$ as $\hat\Lambda = a \Lambda$ and $\hat\Gamma =
a\Lambda$. The affine parameter $\chi$ is related to the redshift $z=1/a
-1$ as
\begin{equation}\label{n9}
 \chi(z) = \int_a^1 dx\, (\Omega_0 x - K x^2 + \lambda_0 x^4)^{-1/2}.
\end{equation}

From Eq. (\ref{n7}) we obtain ${\cal R} = K + \nabla^2 \Phi$ and
${\cal F} = \epsilon^i\epsilon^j \nabla_{\!\!i}\nabla_{\!\!j} \Phi$. We
regard $\delta\!{\cal R}=\nabla^2\Phi$ and $\delta\!{\cal
F}=\epsilon^i\epsilon^j \nabla_{\!\!i}\nabla_{\!\!j} \Phi$ as
perturbations to $\bar{\cal R}=K$ and $\bar{\cal F}=0$, and solve
Eqs. (\ref{n4}) and (\ref{n5}) peturbatively. The zeroth order
solution in an exactly homogenous universe is
\begin{equation}\label{n10}
 \bar\Lambda(\chi) = \bar D(\chi) = \left\{
\begin{array}{ll}
 K^{-1/2} \sin(K^{1/2}\chi), & (K>0) \\
 \chi,                       & (K=0) \\
 (-K)^{-1/2} \sinh[(-K)^{1/2}\chi], & (K<0)
\end{array} \right.
\end{equation}
and $\bar\Gamma(\chi)=0$, and hence the image remains undistorted with only an
isotropic focusing. The differential equation for the first order
quantities $\delta\!\Lambda= \Lambda - \bar\Lambda$ and $\delta\!\Gamma=
\Gamma - \bar\Gamma$ are
\begin{equation}
 \ddot{\delta\!\Lambda} = -(K\,\delta\!\Lambda 
+ \bar D\,\delta\!{\cal R}) \,,\quad \ddot{\delta\!\Gamma} = 
-(K\,\delta\!\Gamma + \bar D\,\delta\!{\cal F}),
\end{equation}
which are integrated with the initial conditions $0= \delta\!\Lambda=
\delta\!\Gamma= \dot{\delta\!\Lambda}= \dot{\delta\!\Gamma}$ as
\begin{equation}\label{n12}
 \delta\!\Lambda(\chi) = -\int_0^\chi d\chi' \, \delta\!{\cal R}(\chi')
 \, \bar D(\chi') \,\bar D(\chi-\chi'),
\end{equation}
\begin{equation}\label{n13}
 \delta\!\Gamma(\chi) = -\int_0^\chi d\chi' \, \delta\!{\cal F}(\chi')
 \, \bar D(\chi') \,\bar D(\chi-\chi').
\end{equation}
Note that $\delta\!\Lambda$ is real, so that there is no rotation 
in the image to this order. Let us define
\begin{equation}
 \kappa = -\delta\!\Lambda/\bar D \,,\quad \gamma = \delta\!\Gamma/\bar D.
\end{equation}
The amplification of the brightness of images relative to that in a
homogenous universe is given by $(\bar D/D)^2=1+2\kappa$ to first order,
so $\kappa$ (convergence) measures the fluctuation in the image
brightness due to the lensing by the density inhomogeneity. Similiarly
$\gamma$ (shear) is a dimensionless measure of the image distortion due
to the lensing.

The sources $\delta\!{\cal R}$ and $\delta\!{\cal F}$ of the image
distortion are stochastic variables which average to zero: $\langle
\delta\!{\cal R}\rangle = \langle \delta\!{\cal F}\rangle =0$. To
evaluate the statistical properties of this distortion, it is
convenient to Fourier-transform the density field at time $\eta$ as
$\tilde\delta(\bm k,\eta)= \int d^3\bm x\, \delta(\bm
x,\eta)\,\exp(-i\bm k\cdot \bm x)$ (valid for $k^2\gg |K|$) and define
the power spectrum $P(k,\eta)$ as $\langle \tilde\delta (\bm k,\eta)\,
\tilde\delta (\bm k',\eta) \rangle = (2\pi)^3 P(k,\eta)\,
\delta_{\rm\scriptscriptstyle D} (\bm k- \bm k')$. Then from Eq.
(\ref{n8}) the two point correlation of $\delta\!{\cal R}$ at two
locations $\bm x$ and $\bm x'$ is
\begin{equation}
 \langle\delta\!{\cal R}(\bm x)\, \delta\!{\cal R}(\bm x') \rangle = 
 \frac94 \Omega_0^2 \int 
  \frac{d^3\bm k}{(2\pi)^3} \,\frac{P(k,\eta)}{a^2(\eta)} \,
  \exp[i\bm k\cdot (\bm x-\bm x')].
\end{equation}
Below we calculate the two point correlations of convergence
$\langle\kappa_{\rm\scriptscriptstyle A}\, \kappa_{\rm\scriptscriptstyle
B} \rangle$ and shear $\langle\gamma_{\rm\scriptscriptstyle A}\,
\gamma_{\rm\scriptscriptstyle B}^* \rangle$ between images A and B at
the same redshifts separated by a small angle vector $\bm\theta$ on the
sky.  Since the integrals in Eqs. (\ref{n12}) and (\ref{n13}) can
be performed along the unperturbed path to first order, $\bm k\cdot
(\bm x-\bm x')$ in the exponent is written as $\bm
k_{\scriptscriptstyle\perp}\cdot \bm\theta \bar D(\chi) +
k_{\scriptscriptstyle\parallel} (\chi-\chi')$ where $\bm
k_{\scriptscriptstyle\perp}$ (or $k_{\scriptscriptstyle\parallel}$) is
the wavenumber perpendicular (or parallel) to the line of sight. When
$\theta\ll 1$ we can approximate $k_{\scriptscriptstyle\parallel} \ll
k_{\scriptscriptstyle\perp}$ and obtain
\begin{equation}\label{n16}
 \langle\delta\!{\cal R}(\bm x)\, \delta\!{\cal R}(\bm x') \rangle 
 = \frac{9\pi}4 \Omega_0^2 \, \delta_{\rm\scriptscriptstyle D}
(\chi-\chi') \int_0^\infty \frac{dk}k\,\frac{\Delta^2(k,\eta)}
{k\, a^2(\eta)} \,J_0[k\theta\bar D(\chi)],
\end{equation}
where $\Delta^2(k,\eta) = k^3\,P(k,\eta)/(2\pi^2)$. A similar
calculation shows that $\langle \delta\!{\cal F}(\bm x)\, \delta\!{\cal
F}^*(\bm x') \rangle = \langle \delta\!{\cal R}(\bm x)\, \delta\!{\cal
R}(\bm x') \rangle$ and $\langle \delta\!{\cal R}(\bm x)\, \delta\!{\cal
F}(\bm x') \rangle =0$. From Eqs. (\ref{n12}), (\ref{n13}) and
(\ref{n16}) there results \footnote{ In conventional units, Eq.
(\ref{n17}) should be divided by $(c/H_0)^4$.}
\begin{eqnarray}
 C(\theta) &=& \langle \kappa_{\rm\scriptscriptstyle A}\, 
\kappa_{\rm\scriptscriptstyle B} \rangle 
 = \langle \gamma_{\rm\scriptscriptstyle A} \,
\gamma_{\rm\scriptscriptstyle B}^* \rangle \nonumber \\
&=& \frac{9\pi}4 \Omega_0^2 \int_0^\chi d\chi'\, 
\frac{\bar D^2(\chi') \bar D^2(\chi-\chi')}{\bar D^2(\chi)} \nonumber \\
&&\times \int_0^\infty \frac{dk}k \,\frac{\Delta^2(k,\chi')}{k\, a^2(\chi')}\, 
J_0[k\theta\bar D(\chi')], \label{n17}
\end{eqnarray}
where $a(\chi')$ is given as Eq. (\ref{n9}).

\begin{figure}[htb]
\epsfxsize=8cm \centerline{\epsfbox{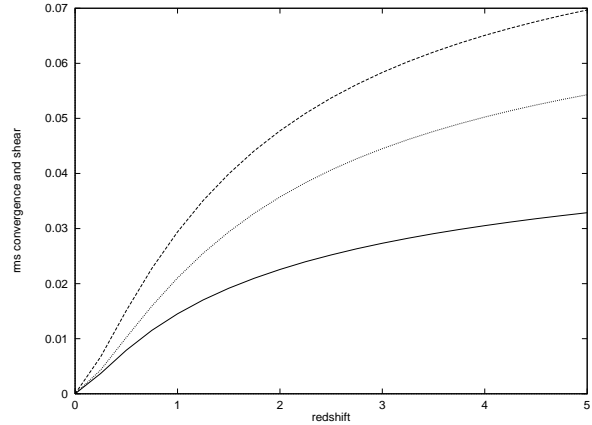}} \caption{Rms convergence
 and shear $C^{1/2}(0)=\langle \kappa^2 \rangle^{1/2} = \langle
 |\gamma|^2 \rangle^{1/2}$ as a function of the source redshift. The solid, dashed
 and dotted curves correspond to $(\Omega_0, \lambda_0, \sigma_8)=
 (1,0,0.5)$, $(0.3,0,1)$, $(0.3,0.7,1)$.} \label{nfig1}
\end{figure}

\begin{figure}[htb]
\epsfxsize=8cm \centerline{\epsfbox{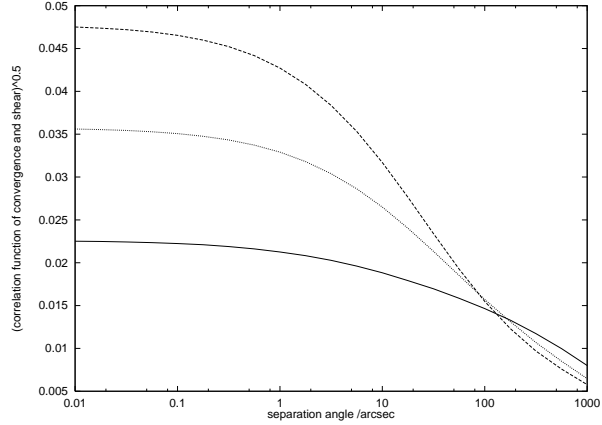}} \caption{Two point
 correlation $C^{1/2}(\theta)=\langle \kappa_{\rm\scriptscriptstyle A}\,
 \kappa_{\rm\scriptscriptstyle B} \rangle^{1/2} = \langle
 \gamma_{\rm\scriptscriptstyle A} \, \gamma_{\rm\scriptscriptstyle B}^*
 \rangle^{1/2}$ of convergence and shear between images A and B at
 redshift $z=2$ as a function of the separation angle $\theta$ in the sky. The
 solid, dashed and dotted curves correspond to $(\Omega_0, \lambda_0,
 \sigma_8)= (1,0,0.5)$, $(0.3,0,1)$, $(0.3,0.7,1)$.} \label{nfig2}
\end{figure}

In Fig.~\ref{nfig1}\ the root-mean-square convergence and shear
$C^{1/2}(0)=\langle \kappa^2 \rangle^{1/2} = \langle |\gamma|^2
\rangle^{1/2}$ is plotted versus the source redshift, and in Fig.
\ref{nfig2} the square root of the two point correlation
$C^{1/2}(\theta)=\langle \kappa_{\rm\scriptscriptstyle A}\,
\kappa_{\rm\scriptscriptstyle B} \rangle^{1/2} = \langle
\gamma_{\rm\scriptscriptstyle A} \, \gamma_{\rm\scriptscriptstyle B}^*
\rangle^{1/2}$ of convergence and shear between images A and B at
redshift $z=2$ is plotted versus the separation angle $\theta$ on the
sky. We use the non-linear power spectrum with the shape parameter
$\Omega_0 h=0.25$, \cite{PEA94} which reproduces well the nearby galaxy
survey data. The solid, dashed and dotted curves in the figures
correspond respectively to the cosmological models with $(\Omega_0,
\lambda_0, \sigma_8)=(1,0,0.5)$, $(0.3,0,1)$, $(0.3,0.7,1)$ (standard,
open and cosmological-constant models), where the normalizations
$\sigma_8$ of the power spectrum on the $8{\rm Mpc}/h$ scale are chosen
such that the observed abundance of low-redshift clusters is consistent
with the prediction of Press--Schechter theory.\cite{KIT97} The figures
represent the result for an infinitesimal source size because we do not
smooth the density field; for a finite source the density field should
be smoothed on the scale of the source size.

Since we are using a power spectrum well-constrained by the
observational data at $z\sim 0$, the differences between the curves in
the figures purely reflect the different behavior of backward light
propagation into the past in these cosmological models. Intuitively, the
rms amplification and shear should be larger in higher density
universes, as explained by the overall factor $\Omega_0^2$ in Eq.
(\ref{n17}). However Fig.~\ref{nfig1} \ shows that this effect is
overwhelmed by the evolutionary effect of the density inhomogeneity: the
growth of the density contrast in the standard (or open) model is
fastest (or slowest) among the three models according to the
gravitational instability picture.\cite{PEE80} Since a fast growth
implies that the inhomogeneity is rapidly erased as one goes to the past,
the lensing effect in the standard (or open) model is smallest (or
largest) in Fig.~\ref{nfig1}. Therefore the detection of cosmic shear
should allow us to probe the growth of the large-scale density
inhomogeneity, when combined with observational constraints for $z\sim
0$. On the other hand, Fig.~\ref{nfig2} \ shows that in the open (or
standard) model the decrease of correlation with increasing angle is
fastest (or slowest). This is due to the cosmological geometry
(Eqs.(\ref{n9}) and (\ref{n10})): for fixed $\theta$ and $z$ the
physical separation $\theta \bar D$ in the open (or standard) model is
largest (or smallest) among the three models. Thus the slope of the
correlation function of cosmic shear, $dC/d\ln\theta$, should be a
measure of the geometry of our universe.

\section{Concluding remarks}

In \S 2 it was shown that the multi-lens-plane method can treat many
ray bundles and realistic galaxy lens models in very useful and
practical ways to
derive statistical behavior of strong and weak lensing owing to the
simplicity of the calculations. The discussion in \S 3 shows that the 
recent progress in
supercomputers has made it possible to use the direct integration methods for 
lensing analyses, and that non-galactic lens objects on galactic scales (if
any) may have more important influences on the cosmological lensing 
(than galactic lenses) on small-angle scales ($< 1$ min) 
depending on the lensing
models. In \S 4 it was shown that perturbative methods based on the CDM
power spectrum are very useful for analyses of weak lensing on
large-angle scales ($\gg 1$ min).  
 
For the lensing correction to the luminosity of small objects such as 
super novae we must treat ray bundles with small separation angles
($\leq 2$ arcsec) for which the deflection due to the small-scale
matter distribution on galactic scales is important. 
>From the analysis in \S 3 it is found that the correction can be
larger by a factor of $\sim 5$ than that due to the perturbative 
methods.\cite{rf:fri}

\section*{Acknowledgements}
K.T. would like to thank Y.~Suto for helpful discussions about
$N$-body simulations. His numerical computations were performed on 
the YITP computer system. \ P.W.P \ is indebted to H.~Martel for his
help in preparing the manuscript, \ to \ T.  Futamase for his 
valuable advice, and to the JSPS for a postdoctoral fellowship.
Her computational works was performed on the University
of Texas High Performance Computing Facility funded through R.~Matzner.



\bigskip

\begin{thebibliography}{99}
\bibitem{rf:gunna} 
J.~E.~Gunn,  \JL{Astrophys.~J.,147,1967,61}.
\bibitem{rf:gunnb} 
J.~E.~Gunn,  \JL{Astrophys.~J.,150,1967,737}.
\bibitem{rf:weina}
S.~Weinberg,  \JL{Astrophys.~J.,208,1976,L1}.
\bibitem{rf:bj}
R.~D.~Blandford and M.~Jaroszy${\rm \acute n}$ski,  \JL{Astrophys.~J.,
246,1981,1}.
\bibitem{SCH92} P.~Schneider, J.~Ehlers and E.~E.~Falco, {\it
Gravitational Lenses} (Springer-Verlag, Berlin, 1992).
\bibitem{rf:bn86}
R.~D.~Blandford and R.~Narayan,  \JL{Astrophys.~J.,310,1986,568}.
\bibitem{rf:sw88a}
P.~Schneider and A.~Weiss, \JL{Astrophys.~J., 327,1988,526}.
\bibitem{rf:sw88b}
P.~Schneider and A.~Weiss, \JL{Astrophys.~J., 330,1988,1}.
\bibitem{rf:jppg}
M.~Jaroszy${\rm \acute n}$ski, C.~Park, B.~Paczy${\rm \acute n}$ski
and J.~R. Gott III, \JL{Astrophys.~J., 365,1990,22}. 
\bibitem{rf:jaro91a}
M.~Jaroszy${\rm \acute n}$ski,  \JL{Mon. Not. R. Astron. Soc.,
249, 1991, 430}.
\bibitem{rf:jaro92b}
M.~Jaroszy${\rm \acute n}$ski,  \JL{Mon. Not. R. Astron. Soc.,
255,1992,655}.
\bibitem{rf:pw89}
B.~Paczy${\rm \acute n}$ski and J.~Wambganss, \JL{Astrophys.~J., 
337,1989,581}.
\bibitem{rf:lp90}
M.~H.~Lee and B.~Paczy${\rm \acute n}$ski, \JL{Astrophys.~J.,
357,1990,32}.
\bibitem{rf:wam95}
J.~Wambsganss, R.~Cen, J.~P.~Ostriker and E.~L.~Turner, \JL{Science, 
268,1995,274}.
\bibitem{rf:wam98}
J.~Wambsganss, R.~Cen and J.~P.~Ostriker, \JL{Astrophys.~J., 494,1998,29}.
\bibitem{rf:prem98}
P.~Premadi, H.~Martel and R.~Matzner, \JL{Astrophys.~J., 493,1998,10}.
\bibitem{rf:sachs61}
R.~K.~Sachs, \JL{Proc. R. Soc. London, A264,1961,309}.
\bibitem{rf:kant69}
R.~Kantowski, \JL{Astrophys.~J., 155,1969,89.}
\bibitem{rf:dr3}
C.~C.~Dyer and R.~C.~Roeder,  \JL{Astrophys.~J.,189,1974, 167.}
\bibitem{rf:kant95}
R.~Kantowski, \JL{Astrophys.~J., 447,1995,35.}
\bibitem{rf:wt90}
K.~Watanabe and K.~Tomita, \JL{Astrophys.~J.,355,1990,1.}
\bibitem{rf:hw98}
D.~E.~Holz and R.~M.~Wald, \PR{D58,1998,063501.}
\bibitem{rf:dyoa83}
C.~C.~Dyer and L.~M.~Oattes, \JL{Astrophys.~J.,326,1988,50.}
\bibitem{rf:holz98}
D.~E.~Holz, \JL{Astrophys.~J., 506, 1998, L1}.
\bibitem{rf:tw89}
K.~Tomita and K.~Watanabe, \PTP{82,1989,563.}
\bibitem{rf:tw90}
K.~Tomita and K.~Watanabe, \PTP{83,1990,467.}
\bibitem{rf:ka88}
A.~Kashlinski, \JL{Astrophys.~J.,331,1988,L1.}
\bibitem{rf:tom88} 
K.~Tomita, \JL{Publ. Astron. Soc. Jpn., 40,1988,751.}
\bibitem{rf:tpr89}
K.~Tomita, \PR{D40,1989, 3821.}
\bibitem{rf:sasaki89}
M.~Sasaki, \JL{Mon. Not. R. Astron. Soc.,
240, 1989, 415.}
\bibitem{rf:tom98a}
K.~Tomita, \PTP{99,1998,97.}
\bibitem{rf:tom98b}
K.~Tomita, Astro-ph/9806003.
\bibitem{rf:tom98c}
K.~Tomita, \PTP{100,1998,79.}
\bibitem{rf:bl91}
A.~Babul and M.~H.~Lee, \JL{Mon. Not. R. Astron. Soc.,
250, 1991, 407.}
\bibitem{rf:bland91}
R.~D.~Blandford, A.~B.~Saust, T.~G.~Brainerd and J.~V.~Villumsen,  
\JL{Astrophys.~J., 251,1991,600.}
\bibitem{rf:vill96}
J.~Villumsen, \JL{Mon. Not. R. Astron. Soc., 281,1996,369.}
\bibitem{rf:bern97}
F.~Bernardeau, L.~Van Waerbeke and Y. Mellier, \JL{Astron. \ Astrophys.,
322,1997, 1}.
\bibitem{rf:nakam97}
T.~T.~Nakamura, \JL{Publ. Astron. Soc. Jpn., 49,1997,151.}
\bibitem{rf:lind90}
V.~E.~Linder, \JL{Mon. Not. R. Astron. Soc., 243,1990,353.}, 
\andvol{243,1990,362.}
\bibitem{rf:mart92}
E. Martinez-Gonzalez, J.~L.~Sanz and J.~Silk, \PR{D46, 1992,4193.}
\bibitem{rf:mart97}
E. Martinez-Gonzalez and J.~L.~Sanz, \JL{Astrophys.~J.,484,1997,1.}
\bibitem{rf:cay93}
L.~Cayon, E. Martinez-Gonzalez and J.~L.~Sanz, \JL{Astrophys.~J.,403,
1993,471.}
\bibitem{rf:selj96}
U.~Seljak, \JL{Astrophys.~J.,463,1996,1.}
\bibitem{rf:jain97}
B.~Jain and U.~Seljak, \JL{Astrophys.~J.,484,1997,560.}
\bibitem{rf:HE81}
R.~W.~Hockney and J.~W.~Eastwood, {\it Computer Simulation using Particles}
(New York, McGraw-Hill, 1981)
\bibitem{rf:MPM98}
H.~Martel, P.~Premadi and R.~Matzner, \JL{Astrophys.~J.,497,1998, 512}.
\bibitem{rf:EEP88}
G.~Efstathiou, R.~S.~Ellis and B.~A.~Peterson, \JL{Mon. Not. R. Astron. Soc.,
232,1988, 431}.
\bibitem{rf:BW97}
E.~F.~Bunn and M.~White, \JL{Astrophys.~J., 480, 1997, 6}.
\bibitem{rf:su}
Y.~Suto, \PTP{90,1993,1173}.
\bibitem{rf:bert}
E.~Bertschinger, Astro-ph/9506070.
\bibitem{rf:perl}
S.~Perlmutter et al., \JL{Astrophys.~J.,483,1997,565.}
\bibitem{rf:riess}
A.~G.~Riess et al.,  \JL{Astron.~J.,116,1998,1009.} 
\bibitem{rf:fort}
B.~Ford, Y.~Miller, M.~Dantel-Fort, H.~Bonnet and J.-P. Kneib, 
\JL{Astron. \ Astrophys.,310,1996, 705.}
\bibitem{rf:sch}
P.~Schneider, L.~van Waerkebe, Y.~Miller, B.~Jain, S.~Seits and
B.~Ford, \JL{Astron. \ Astrophys.,333,1998, 767.}
\bibitem{MIS73} C.~W.~Misner, K.~S.~Thorne and J.~A.~Wheeler, {\it
 Gravitation} (Freeman, San Francisco, 1973).
\bibitem{SAS87} M.~Sasaki, \JL{Mon.~Not.~R.~Astron.~Soc., 228,1987,653}.
\bibitem{FUT89} T.~Futamase and M.~Sasaki, \PR{D40,1989,2502}.
\bibitem{WAT90} T.~Watanabe and M.~Sasaki, \JL{Publ.~Astron.~Soc.~Jpn., 
 42,1990,L33}.
\bibitem{MIR91} J.~Miralda-Escud\'e, \JL{Astrophys.~J.,380,1991,1}.
\bibitem{KAI92} N.~Kaiser, \JL{Astrophys.~J.,388,1992,272}.
\bibitem{SEI94} S.~Seitz and P.~Schneider, \JL{Astron.~Astrophys.,
 287,1994,349}.
\bibitem{BAR96} R.~Bar-Kana, \JL{Astrophys. J., 468,1996,17}. 
\bibitem{SAS93} M.~Sasaki, \PTP{90,1993,753.}
\bibitem{PEA94} J.~A.~Peacock and S.~J.~Dodds,
 \JL{Mon.~Not.~R.~Astron.~Soc., 267,1994,1020.}
\bibitem{KIT97} T.~Kitayama and Y.~Suto, \JL{Astrophys.~J.,490,1997,557}.
\bibitem{PEE80} P.~J.~E.~Peebles, {\it The Large-Scale Structure of the
 Universe} (Princeton University Press, Princeton, 1980).
\bibitem{rf:fri} J.~A. Frieman, \JL{Comments Astrophys., 18,1997,323.}


\end{thebibliography}
\end{document}